\pdfoutput=1
\documentclass[12pt,a4paper]{article}

\usepackage{ifthen} 
\newboolean{pdflatex}
\setboolean{pdflatex}{true} 

\newboolean{articletitles}
\setboolean{articletitles}{true} 

\newboolean{uprightparticles}
\setboolean{uprightparticles}{false} 

\def\paperauthors{LHCb collaboration} 
\def\paperasciititle
{Measurement of Lambda_b0 and Lambda_c+ decay parameters} 
\def\papertitle{Measurement of \Lb, \Lc and \Lz decay parameters using  $\Lb\to\Lc h^-$ decays} 
\def\papercopyright{\the\year\ CERN for the benefit of the LHCb collaboration} 
\def\paperlicence{CC BY 4.0 licence}
\def\paperlicenceurl{https://creativecommons.org/licenses/by/4.0/}

\usepackage[top=1in, bottom=1.25in, left=1in, right=1in]{geometry}

\usepackage{microtype}
\usepackage{lineno}  
\usepackage{xspace} 
\usepackage{caption} 

\usepackage{graphicx}  
\usepackage{color}
\usepackage{colortbl}
\graphicspath{{./figs/}} 

\usepackage{amsmath} 
\usepackage{amssymb}
\usepackage{amsfonts}
\usepackage{upgreek} 

\newcommand*\patchAmsMathEnvironmentForLineno[1]{%
\expandafter\let\csname old#1\expandafter\endcsname\csname #1\endcsname
\expandafter\let\csname oldend#1\expandafter\endcsname\csname
end#1\endcsname
 \renewenvironment{#1}%
   {\linenomath\csname old#1\endcsname}%
   {\csname oldend#1\endcsname\endlinenomath}%
}
\newcommand*\patchBothAmsMathEnvironmentsForLineno[1]{%
  \patchAmsMathEnvironmentForLineno{#1}%
  \patchAmsMathEnvironmentForLineno{#1*}%
}
\AtBeginDocument{%
\patchBothAmsMathEnvironmentsForLineno{equation}%
\patchBothAmsMathEnvironmentsForLineno{align}%
\patchBothAmsMathEnvironmentsForLineno{flalign}%
\patchBothAmsMathEnvironmentsForLineno{alignat}%
\patchBothAmsMathEnvironmentsForLineno{gather}%
\patchBothAmsMathEnvironmentsForLineno{multline}%
\patchBothAmsMathEnvironmentsForLineno{eqnarray}%
}

\usepackage{hyperxmp}

\usepackage[pdftex,
            pdfauthor={\paperauthors},
            pdftitle={\paperasciititle},
            pdfcopyright={Copyright (C) \papercopyright},
            pdflicenseurl={\paperlicenceurl}]{hyperref}

\usepackage[colorinlistoftodos,textsize=scriptsize]{todonotes}

\usepackage[bottom,flushmargin,hang,multiple]{footmisc}

\usepackage[all]{hypcap} 

\usepackage{xspace} 
\usepackage{upgreek}

\def\lhcb   {\mbox{LHCb}\xspace}

\def\belle  {\mbox{Belle}\xspace}

\def\besiii {\mbox{BESIII}\xspace}

\def\lhc    {\mbox{LHC}\xspace}

\def\MagUp {\mbox{\em Mag\kern -0.05em Up}\xspace}

\ifthenelse{\boolean{uprightparticles}}%
{

 \def\Ppi         {\ensuremath{\uppi}\xspace}

 \def\Ppsi        {\ensuremath{\uppsi}\xspace}

 \def\PDelta      {\ensuremath{\Delta}\xspace}                 
 \def\PXi         {\ensuremath{\Xi}\xspace}                 
 \def\PLambda     {\ensuremath{\Lambda}\xspace}                 
 \def\PSigma      {\ensuremath{\Sigma}\xspace}                 
 \def\POmega      {\ensuremath{\Omega}\xspace}                 
 \def\PUpsilon    {\ensuremath{\Upsilon}\xspace}
 \let\oldPi\Pi
 \def\PPi         {\ensuremath{\oldPi}\xspace}

 \def\PB      {\ensuremath{\mathrm{B}}\xspace}                 
 \def\PD      {\ensuremath{\mathrm{D}}\xspace}                 
 \def\PJ      {\ensuremath{\mathrm{J}}\xspace}                 
 \def\PK      {\ensuremath{\mathrm{K}}\xspace}                 
 \def\Pb      {\ensuremath{\mathrm{b}}\xspace}                 
 \def\Pc      {\ensuremath{\mathrm{c}}\xspace}

 \def\Pp      {\ensuremath{\mathrm{p}}\xspace}                 

 \def\Ps      {\ensuremath{\mathrm{s}}\xspace}

 \def\thebaroffset{0.0em}
}
{

 \def\Ppi         {\ensuremath{\pi}\xspace}

 \def\Ppsi        {\ensuremath{\psi}\xspace}                 
                  
 \mathchardef\PDelta="7101
 \mathchardef\PXi="7104
 \mathchardef\PLambda="7103
 \mathchardef\PSigma="7106
 \mathchardef\POmega="710A
 \mathchardef\PUpsilon="7107
 \mathchardef\PPi="7105
 \def\PB      {\ensuremath{B}\xspace}                 
 \def\PD      {\ensuremath{D}\xspace}                 
 \def\PJ      {\ensuremath{J}\xspace}                 
 \def\PK      {\ensuremath{K}\xspace}                 
 \def\Pb      {\ensuremath{b}\xspace}                 
 \def\Pc      {\ensuremath{c}\xspace}

 \def\Pp      {\ensuremath{p}\xspace}                 

 \def\Ps      {\ensuremath{s}\xspace}

 \def\thebaroffset{0.18em}
}
\newcommand{\offsetoverline}[2][\thebaroffset]{\kern #1\overline{\kern -#1 #2}}%

\makeatletter
\ifcase \@ptsize \relax
  \newcommand{\miniscule}{\@setfontsize\miniscule{4}{5}}
\or
  \newcommand{\miniscule}{\@setfontsize\miniscule{5}{6}}
\or
  \newcommand{\miniscule}{\@setfontsize\miniscule{5}{6}}
\fi
\makeatother

\DeclareRobustCommand{\optbar}[1]{\shortstack{{\miniscule (\rule[.5ex]{1.25em}{.18mm})}
  \\ [-.7ex] $#1$}}

\def\squark    {{\ensuremath{\Ps}}\xspace}

\def\cquark    {{\ensuremath{\Pc}}\xspace}

\def\bquark    {{\ensuremath{\Pb}}\xspace}

\def\pion   {{\ensuremath{\Ppi}}\xspace}

\def\pip    {{\ensuremath{\pion^+}}\xspace}
\def\pim    {{\ensuremath{\pion^-}}\xspace}

\def\kaon    {{\ensuremath{\PK}}\xspace}

\def\KorKbar {\kern \thebaroffset\optbar{\kern -\thebaroffset \PK}{}\xspace}

\def\Kp      {{\ensuremath{\kaon^+}}\xspace}
\def\Km      {{\ensuremath{\kaon^-}}\xspace}

\def\KS      {{\ensuremath{\kaon^0_{\mathrm{S}}}}\xspace}

\def\D       {{\ensuremath{\PD}}\xspace}

\def\DorDbar {\kern \thebaroffset\optbar{\kern -\thebaroffset \PD}\xspace}
\def\Dz      {{\ensuremath{\D^0}}\xspace}

\def\Dp      {{\ensuremath{\D^+}}\xspace}
\def\Dm      {{\ensuremath{\D^-}}\xspace}

\def\DpDm    {\ensuremath{\Dp {\kern -0.16em \Dm}}\xspace}

\def\B       {{\ensuremath{\PB}}\xspace}

\def\BorBbar {\kern \thebaroffset\optbar{\kern -\thebaroffset \PB}\xspace}

\def\Bd      {{\ensuremath{\B^0}}\xspace}

\def\BdorBdbar {\kern \thebaroffset\optbar{\kern -\thebaroffset \Bd}\xspace}

\def\Bs      {{\ensuremath{\B^0_\squark}}\xspace}

\def\BsorBsbar {\kern \thebaroffset\optbar{\kern -\thebaroffset \Bs}\xspace}

\def\jpsi     {{\ensuremath{{\PJ\mskip -3mu/\mskip -2mu\Ppsi}}}\xspace}

\def\Y#1S{\ensuremath{\PUpsilon{(#1S)}}\xspace}

\def\proton      {{\ensuremath{\Pp}}\xspace}
\def\antiproton  {{\ensuremath{\overline \proton}}\xspace}

\def\Lz          {{\ensuremath{\PLambda}}\xspace}
\def\Lbar        {{\ensuremath{\offsetoverline{\PLambda}}}\xspace}
\def\LorLbar     {\kern \thebaroffset\optbar{\kern -\thebaroffset \PLambda}\xspace}

\def\Sigmares    {{\ensuremath{\PSigma}}\xspace}
\def\Sigmaz      {{\ensuremath{\Sigmares{}^0}}\xspace}

\def\Lc          {{\ensuremath{\Lz^+_\cquark}}\xspace}

\def\Lb           {{\ensuremath{\Lz^0_\bquark}}\xspace}
\def\Lbbar        {{\ensuremath{\Lbar{}^0_\bquark}}\xspace}

\def\to                 {\ensuremath{\rightarrow}\xspace}

\def\CP                {{\ensuremath{C\!P}}\xspace}

\def\AT#1     {\ensuremath{A_{\mathrm{T}}^{#1}}\xspace}           

\def\C#1      {\ensuremath{\mathcal{C}_{#1}}\xspace}                       
\def\Cp#1     {\ensuremath{\mathcal{C}_{#1}^{'}}\xspace}                    
\def\Ceff#1   {\ensuremath{\mathcal{C}_{#1}^{\mathrm{(eff)}}}\xspace}        
\def\Cpeff#1  {\ensuremath{\mathcal{C}_{#1}^{'\mathrm{(eff)}}}\xspace}       
\def\Ope#1    {\ensuremath{\mathcal{O}_{#1}}\xspace}                       
\def\Opep#1   {\ensuremath{\mathcal{O}_{#1}^{'}}\xspace}                    

\newcommand{\aunit}[1]{\ensuremath{\text{\,#1}}}       

\newcommand{\tev}{\aunit{Te\kern -0.1em V}\xspace}
\newcommand{\gev}{\aunit{Ge\kern -0.1em V}\xspace}
\newcommand{\mev}{\aunit{Me\kern -0.1em V}\xspace}
\newcommand{\kev}{\aunit{ke\kern -0.1em V}\xspace}
\newcommand{\ev}{\aunit{e\kern -0.1em V}\xspace}
 
\newcommand{\mevc}{\ensuremath{\aunit{Me\kern -0.1em V\!/}c}\xspace}
\newcommand{\gevc}{\ensuremath{\aunit{Ge\kern -0.1em V\!/}c}\xspace}
\newcommand{\mevcc}{\ensuremath{\aunit{Me\kern -0.1em V\!/}c^2}\xspace}
\newcommand{\gevcc}{\ensuremath{\aunit{Ge\kern -0.1em V\!/}c^2}\xspace}

\def\fb   {\ensuremath{\aunit{fb}}\xspace}
\def\invfb   {\ensuremath{\fb^{-1}}\xspace}

\newcommand{\chisq}{\ensuremath{\chi^2}\xspace}

\def\gsim{{~\raise.15em\hbox{$>$}\kern-.85em
          \lower.35em\hbox{$\sim$}~}\xspace}
\def\lsim{{~\raise.15em\hbox{$<$}\kern-.85em
          \lower.35em\hbox{$\sim$}~}\xspace}

\def\sPlot{\mbox{\em sPlot}\xspace}

\def\sqs   {\ensuremath{\protect\sqrt{s}}\xspace}

\def\rad{\aunit{rad}\xspace}

\def\tell1  {TELL1\xspace}
\def\ukl1   {UKL1\xspace}

\newcommand{\lhcborcid}[1]{\href{https://orcid.org/#1}{\hspace*{0.1em}\raisebox{-0.45ex}{\includegraphics[width=1em]{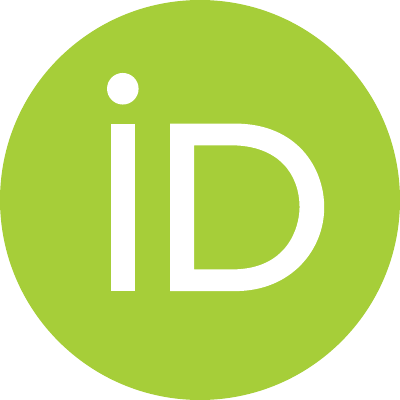}}}}

\usepackage{cite} 
\usepackage{mciteplus}

\def\LbToLcpiToLzpi{{\ensuremath{\Lb\to\Lc(\to\Lz\pip)\pim}}\xspace}
\def\LbToLcpiToLzK{{\ensuremath{\Lb\to\Lc(\to\Lz\Kp)\pim}}\xspace}
\def\LbToLcKToLzpi{{\ensuremath{\Lb\to\Lc(\to\Lz\pip)\Km}}\xspace}

\def\LbToLcpiTopKS{{\ensuremath{\Lb\to\Lc(\to\proton\KS)\pim}}\xspace}
\def\LbToLcKTopKS{{\ensuremath{\Lb\to\Lc(\to\proton\KS)\Km}}\xspace}
\def\LbToLchToLh{\mbox{{\ensuremath{\Lb\to\Lc(\to\Lz(\to\proton\pim) h_1^{+})h_2^{-}}}}\xspace}

\def\LbToLcpiToLzh{\mbox{{\ensuremath{\Lb\to\Lc(\to\Lz h^{+})\pim}}}\xspace}
\def\LbToLchTopKS{\mbox{{\ensuremath{\Lb\to\Lc(\to\proton\KS) h^{-}}}}\xspace}
\def\LbToLchhTopKS{\mbox{{\ensuremath{\Lb\to\Lc(\to\proton\KS) h_{2}^{-}}}}\xspace}

\def\LbnLchLambdahdecay{{\ensuremath{\Lb\to(\Lc\to\Lz h_1^{+})h_2^{-}}}\xspace}

\def\LbnLchKspdecay{{\ensuremath{\Lb\to(\Lc\to\proton\KS)h^{-}}}\xspace}

\def\alphaB{\ensuremath{\alpha_{\Lb}}\xspace}
\def\betaB{\ensuremath{\beta_{\Lb}}\xspace}
\def\gammaB{\ensuremath{\gamma_{\Lb}}\xspace}

\def\alphaC{\ensuremath{\alpha_{\Lc}}\xspace}
\def\betaC{\ensuremath{\beta_{\Lc}}\xspace}
\def\gammaC{\ensuremath{\gamma_{\Lc}}\xspace}
\def\alphaCh{\ensuremath{\alpha_{\Lc}}\xspace}
\def\betaCh{\ensuremath{\beta_{\Lc}}\xspace}
\def\gammaCh{\ensuremath{\gamma_{\Lc}}\xspace}
\def\DeltaCh{\ensuremath{\Delta_{\Lc}}\xspace}

\def\alphaCp{\ensuremath{\alpha_{\Lc}}\xspace}

\def\alphaS{\ensuremath{\alpha_{\Lz}}\xspace}

\def\myLcbar          {{\ensuremath{\Lz^-_\cquark}}\xspace}

\def\aBpi{\ensuremath{\alpha_{\Lb}^{\Lc\pim}}\xspace}
\def\aBk{\ensuremath{\alpha_{\Lb}^{\Lc\Km}}\xspace}
\def\aCpi{\ensuremath{\alpha_{\Lc}^{\Lz\pip}}\xspace}
\def\bCpi{\ensuremath{\beta_{\Lc}^{\Lz\pip}}\xspace}
\def\gCpi{\ensuremath{\gamma_{\Lc}^{\Lz\pip}}\xspace}
\def\dCpi{\ensuremath{\Delta_{\Lc}^{\Lz\pip}}\xspace}
\def\aCk{\ensuremath{\alpha_{\Lc}^{\Lz\Kp}}\xspace}
\def\bCk{\ensuremath{\beta_{\Lc}^{\Lz\Kp}}\xspace}
\def\gCk{\ensuremath{\gamma_{\Lc}^{\Lz\Kp}}\xspace}
\def\dCk{\ensuremath{\Delta_{\Lc}^{\Lz\Kp}}\xspace}
\def\aCp{\ensuremath{\alpha_{\Lc}^{\proton\KS}}\xspace}
\def\aS{\ensuremath{\alpha_{\Lz}^{\proton\pim}}\xspace}

\def\baBpi{\ensuremath{\overline{\alpha}_{\Lbbar}^{\myLcbar\pip}}\xspace}
\def\baBk{\ensuremath{\overline{\alpha}_{\Lbbar}^{\myLcbar\Kp}}\xspace}
\def\baCpi{\ensuremath{\overline{\alpha}_{\myLcbar}^{\Lbar\pim}}\xspace}
\def\bbCpi{\ensuremath{\overline{\beta}_{\myLcbar}^{\Lbar\pim}}\xspace}
\def\bgCpi{\ensuremath{\overline{\gamma}_{\myLcbar}^{\Lbar\pim}}\xspace}
\def\bdCpi{\ensuremath{\overline{\Delta}_{\myLcbar}^{\Lbar\pim}}\xspace}
\def\baCk{\ensuremath{\overline{\alpha}_{\myLcbar}^{\Lbar\Km}}\xspace}
\def\bbCk{\ensuremath{\overline{\beta}_{\myLcbar}^{\Lbar\Km}}\xspace}
\def\bgCk{\ensuremath{\overline{\gamma}_{\myLcbar}^{\Lbar\Km}}\xspace}
\def\bdCk{\ensuremath{\overline{\Delta}_{\myLcbar}^{\Lbar\Km}}\xspace}
\def\baCp{\ensuremath{\overline{\alpha}_{\myLcbar}^{\antiproton\KS}}\xspace}
\def\baS{\ensuremath{\overline{\alpha}_{\Lbar}^{\antiproton\pip}}\xspace}

\def\avgBpi{\ensuremath{\langle\alpha_{\Lb}^{\Lc\pim}\rangle}\xspace}
\def\avgBk{\ensuremath{\langle\alpha_{\Lb}^{\Lc\Km}\rangle}\xspace}
\def\avgCpi{\ensuremath{\langle\alpha_{\Lc}^{\Lz\pip}\rangle}\xspace}
\def\avgCk{\ensuremath{\langle\alpha_{\Lc}^{\Lz\Kp}\rangle}\xspace}
\def\avgCp{\ensuremath{\langle\alpha_{\Lc}^{\proton\KS}\rangle}\xspace}
\def\avgS{\ensuremath{\langle\alpha_{\Lz}^{\proton\pim}\rangle}\xspace}

\def\acpBpi{\ensuremath{{A_\alpha}_{\Lb}^{\Lc\pim}}\xspace}
\def\acpBk{\ensuremath{{A_\alpha}_{\Lb}^{\Lc\Km}}\xspace}
\def\acpCpi{\ensuremath{{A_\alpha}_{\Lc}^{\Lz\pip}}\xspace}
\def\acpCk{\ensuremath{{A_\alpha}_{\Lc}^{\Lz\Kp}}\xspace}
\def\acpCp{\ensuremath{{A_\alpha}_{\Lc}^{\proton\KS}}\xspace}
\def\acpS{\ensuremath{{A_\alpha}_{\Lz}^{\proton\pim}}\xspace}

\def\rCpi{\ensuremath{{R_\beta}_{\Lc}^{\Lz\pip}}\xspace}
\def\rCk{\ensuremath{{R_\beta}_{\Lc}^{\Lz\Kp}}\xspace}

\def\rrCpi{\ensuremath{{R'_\beta}_{\Lc}^{\Lz\pip}}\xspace}
\def\rrCk{\ensuremath{{R'_\beta}_{\Lc}^{\Lz\Kp}}\xspace}

\newcommand{\xx}{\ensuremath{\kern 0.5em }}

\def\sPlot{\mbox{\em sPlot}\xspace}

\usepackage{multirow} 
\usepackage{booktabs} 
\usepackage{rotating}

\usepackage{longtable} 
\usepackage{multirow}
\usepackage{pdflscape}

\begin{document}

\renewcommand{\thefootnote}{\fnsymbol{footnote}}
\setcounter{footnote}{1}


\begin{titlepage}
\pagenumbering{roman}

\vspace*{-1.5cm}
\centerline{\large EUROPEAN ORGANIZATION FOR NUCLEAR RESEARCH (CERN)}
\vspace*{1.5cm}
\noindent
\begin{tabular*}{\linewidth}{lc@{\extracolsep{\fill}}r@{\extracolsep{0pt}}}
\ifthenelse{\boolean{pdflatex}}
{\vspace*{-1.5cm}\mbox{\!\!\!\includegraphics[width=.14\textwidth]{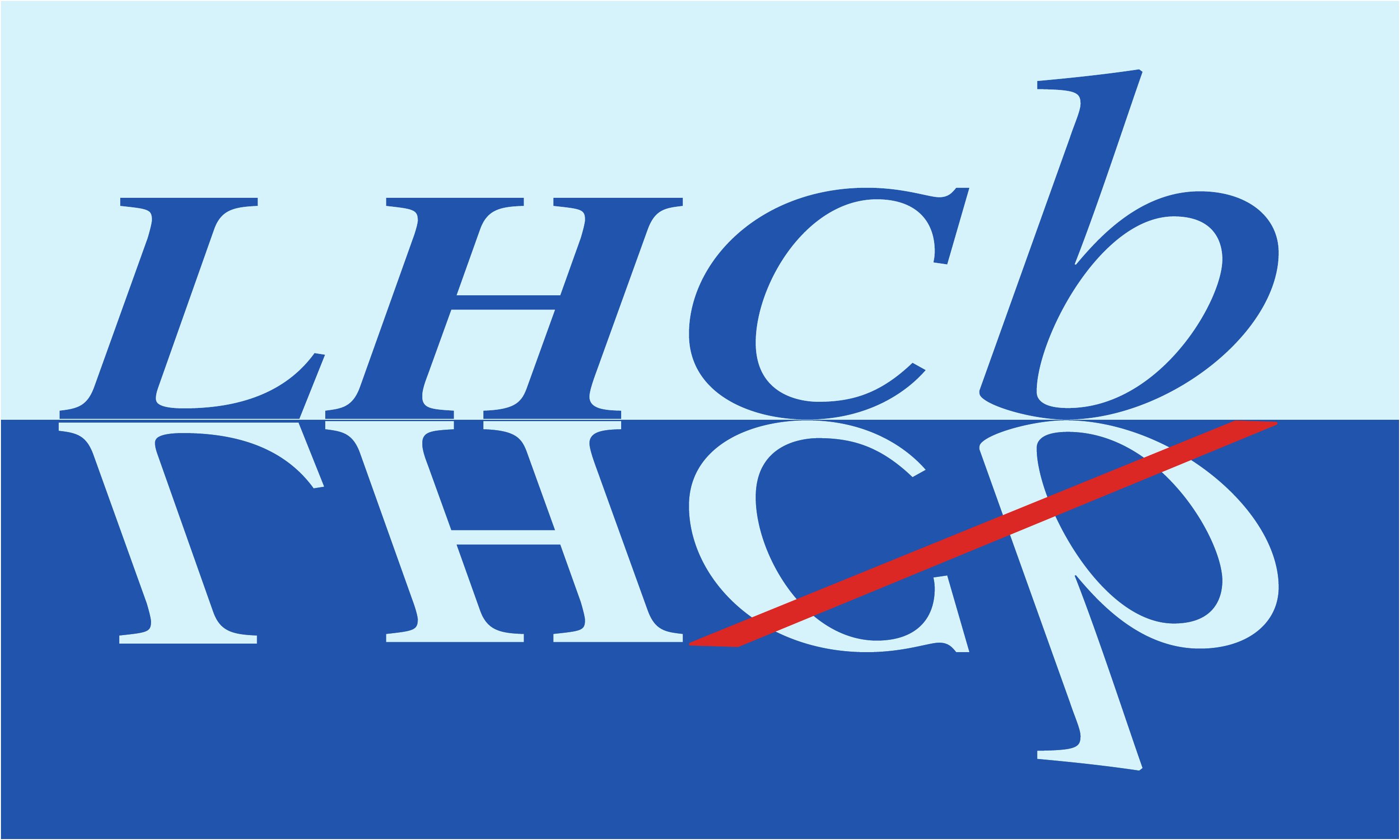}} & &}%
{\vspace*{-1.2cm}\mbox{\!\!\!\includegraphics[width=.12\textwidth]{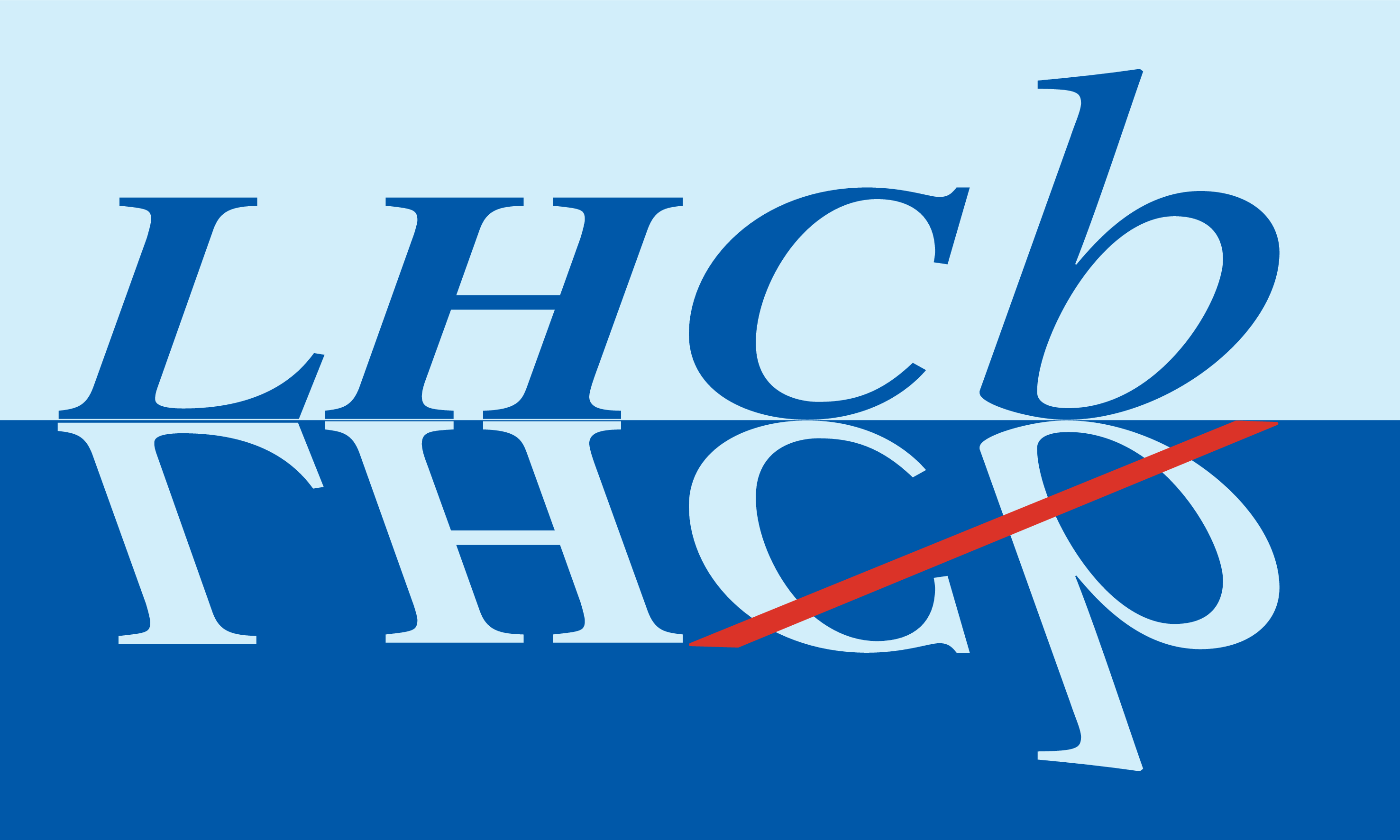}} & &}%
\\
 & & CERN-EP-2024-200 \\  
 & & LHCb-PAPER-2024-017 \\  
 & & January 7, 2025 \\
 & & \\
\end{tabular*}

\vspace*{1.0cm}

{\normalfont\bfseries\boldmath\huge
\begin{center}
  \papertitle 
\end{center}
}

\vspace*{1.0cm}

\begin{center}
\paperauthors\footnote{Authors are listed at the end of this paper.}
\end{center}

\vspace{\fill}

\begin{abstract}
  \noindent

A comprehensive study of the angular distributions in the bottom-baryon decays $\Lb\to\Lc h^-(h=\pi, K)$, followed by $\Lc\to\Lz h^+$ with $\Lz\to \proton \pim$ or $\Lc\to\proton\KS$ decays, is performed 
using a data sample of proton-proton collisions corresponding to an integrated luminosity of $9\invfb$ collected by the LHCb experiment at center-of-mass energies of 7, 8 and 13 \tev.
The decay parameters and the associated charge-parity (\CP) asymmetries are measured,
with no significant \CP violation observed.
For the first time, the $\Lb \to \Lc h^-$ decay parameters are measured.
The most precise measurements of the decay parameters $\alpha, \beta$ and $\gamma$ are obtained for $\Lc$ decays and 
an independent measurement of the decay parameters for the strange-baryon $\Lz$ decay is provided.
The results deepen our understanding of weak decay dynamics in baryon decays.


\end{abstract}

\vspace*{1.0cm}

\begin{center}
  Published in
  Physical~Review~Letters~133~(2024)~261804
\end{center}

\vspace{\fill}

{\footnotesize 
\centerline{\copyright~\papercopyright. \href{\paperlicenceurl}{\paperlicence}.}}
\vspace*{2mm}

\end{titlepage}


\newpage
\setcounter{page}{2}
\mbox{~}
%
%
%
%


\renewcommand{\thefootnote}{\arabic{footnote}}
\setcounter{footnote}{0}


\cleardoublepage


\pagestyle{plain} 
\setcounter{page}{1}
\pagenumbering{arabic}



Hadronic weak decays of baryons provide an excellent platform for 
studying baryon decay dynamics and the origin of the asymmetry between matter and antimatter~\cite{PhysRevD.34.833,Sakharov:1967dj,2022Symm}.
Among them, the  decay of a spin-half baryon to a spin-half baryon and a pseudoscalar meson is of special interest. 
For this type of decay, three decay parameters, first proposed by Lee and Yang to search for parity violation~\cite{PhysRev.108.1645}, can be defined as
\begin{equation}
\alpha\equiv\frac{2\Re(s^*p)}{|s|^2+|p|^2}, \;\;\beta\equiv\frac{2\Im(s^*p)}{|s|^2+|p|^2},\;\; \gamma\equiv\frac{|s|^2-|p|^2}{|s|^2+|p|^2} \,,
\end{equation}
satisfying $\alpha^2+\beta^2+\gamma^2=1$,   where $s$ and $p$ denote the parity-violating S-wave and parity-conserving P-wave amplitudes, respectively.
The interference between the two amplitudes may
generate differences between the differential decay rates of baryons and antibaryons, allowing  \CP-violation phenomena to be probed via  angular analyses~\cite{Dai:2023zms}.
The amount of \CP violation  can be quantified by the asymmetries
$A_{\alpha}=(\alpha+\bar{\alpha})/(\alpha-\bar{\alpha})$ and
$R_{\beta}=(\beta+\bar{\beta})/(\alpha-\bar{\alpha})$, 
where $\bar{\alpha}$ and $\bar{\beta}$ denote the decay parameters of the antibaryons, and should have signs opposite to their baryonic counterparts.
At leading order, these \CP asymmetries are related to the weak and strong phase differences between the S- and P-wave amplitudes, $\Delta{\phi}$ and $\Delta{\delta}$, via the relations 
$A_{\alpha}=-\tan\Delta{\delta}\tan\Delta{\phi}$ and
$R_{\beta}=\tan\Delta{\phi}$~\cite{PhysRevD.34.833}.

Many phenomenological models have been used to calculate baryon decay parameters.
For some two-body beauty-baryon decays, factorization is assumed to hold in model calculations~\cite{PhysRevD.77.014020, Leibovich:2003tw,PhysRevD.61.114002,Mannel:1992ti,PhysRevD.102.034033,PhysRevD.99.054020,Mohanta1999,PhysRevD.100.034025,Ke:2019smy,PhysRevD.105.073005}, which predict that  \mbox{$\alpha  \approx -1$}, consistent with the $V-A$ nature of the weak current and maximal parity violation.
For charm-baryon decays, model calculations are complicated by the presence of nonfactorizable contributions and often do not agree with each other~\cite{PhysRevD.55.7067,PhysRevD.40.1513,PhysRevD.48.4188,PhysRevD.46.270,PhysRevD.49.3417,PhysRevD.50.5787,PhysRevD.101.014011,Korner:1992wi,Sharma:1998rd,PhysRevD.97.074028,PhysRevD.57.5632,Geng:2019xbo}.
For strange-baryon decays, nonfactorizable  contributions may dominate, making  theoretical calculations even more challenging~\cite{PhysRevD.34.833}.

Decay parameters have been measured for several  hyperon and charm-baryon decays~\cite{ParticleDataGroup:2024cfk},
while beauty decays are much less explored.
The $\alpha$ parameter of the $\Lz\to\proton\pim$ decay was recently updated by the \besiii~\cite{BESIII:2018cnd,PhysRevLett.129.131801} and CLAS~\cite{Ireland:2019uja} collaborations, which resulted in a significantly larger value compared to the previous world average~\cite{PDG2018}.
The $\alpha$ parameters of several \Lc decays were precisely measured by the FOCUS~\cite{FOCUS:2005vxq}, \besiii~\cite{BESIII:2019odb} and \belle~\cite{Belle:2022uod} collaborations,
while the precision of the $\beta$ and $\gamma$ measurements  is still very limited~\cite{BESIII:2019odb,PhysRevLett.132.031801}.
To date, there is no 
decay parameter measurement  for any  $\Lb$  decay to a baryon and a pseudoscalar meson, despite the observation of many such  decay modes.
The decay parameter of the $\Lb\rightarrow \jpsi\Lz$ decay  was measured in proton-proton ($pp$) collisions at the \lhc~\cite{LHCB-PAPER-2012-057,LHCb-PAPER-2020-005,ATLAS:2014swk,CMS:2018wjk},
together with the $\Lb$ polarization, which is  found to be consistent with zero.
Moreover, the photon polarization of the $\Lb\to\Lz\gamma$ decay was measured by LHCb~\cite{LHCb-PAPER-2021-030}, suggesting the dominance of left-handed photons.

In this Letter, the decay parameters and \CP asymmetries of \mbox{$\Lb\to\Lc \pim$} and  \mbox{$\Lb\to\Lc \Km$}  decays are measured through an angular analysis.
Three  $\Lc$ decays are analyzed:
$\Lc\to p\KS$, 
$\Lc\to\Lz\pip$
and $\Lc\to\Lz\Kp$ with the subsequent decays $\Lz\to\proton\pim$ and $\KS \to \pi^+\pi^-$.
The decay parameters and associated \CP asymmetries  of the 
$\Lb$, $\Lc$ and $\Lz$ decays are  determined simultaneously.
The analysis is performed using data from $pp$ collisions at center-of-mass energies of $\sqs=7$, $8$  and $13\tev$, 
corresponding to an integrated luminosity of $9\invfb$ collected with the~\lhcb detector.  Inclusion of charge-conjugate processes is implied, unless otherwise stated.

The LHCb detector, designed for the study of particles containing \bquark\ or \cquark\ quarks, is a single-arm forward spectrometer covering the pseudorapidity range $2 < \eta < 5$, described in detail in Refs.~\cite{LHCb-DP-2008-001,LHCb-DP-2014-002}. 
The online event selection for \Lb decays is performed by a trigger~\cite{LHCb-DP-2012-004}, which consists of a hardware stage followed by a  software stage~\cite{LHCb-TDR-016, LHCb-PROC-2015-018, BBDT, LHCb-DP-2019-001}.
The hardware trigger is based on information from the calorimeter and muon systems.
The software trigger requires 
a secondary vertex with a significant displacement from any primary vertex (PV).

Simulated samples of $\Lb$ decays are produced to optimize event selection, study potential backgrounds and model the detector acceptance.
These samples are generated using the software described in Refs.~\cite{Sjostrand:2007gs,*Sjostrand:2006za,LHCb-PROC-2010-056,Lange:2001uf,davidson2015photos,Allison:2006ve, *Agostinelli:2002hh, LHCb-PROC-2011-006}. 
The products of each decay in the $\Lb$ cascades are distributed uniformly in the allowed phase space.

In the offline selection, all tracks in the final state are required to have a  large transverse momentum and be inconsistent with being directly produced from any PV\@.
The \Lz and \KS candidates are reconstructed using $\Lz\to\proton\pim$ and $\KS\to\pim\pip$ decays, 
where the final-state tracks are required to form a vertex with a good fit quality that is significantly displaced from any PV\@, 
and their invariant mass is consistent with the known value~\cite{ParticleDataGroup:2024cfk}.
The \Lz (\KS) candidate is combined with a kaon/pion (proton) track to form the $\Lc$ candidate.
The \Lc
invariant mass is required to be within $\pm26\,(20)\mevcc$ of the known value~\cite{ParticleDataGroup:2024cfk} for the $\Lc\to\proton\KS$ and $\Lc\to\Lz\pip$ ($\Lc\to\Lz\Kp$) decays. The smaller mass region for the $\Lc\to\Lz\Kp$ decay is used to suppress the $\Lc\to\Sigmaz(\to\Lz\gamma)\pip$ background, where the photon is not reconstructed.
The \Lb candidate is formed by combining a \Lc candidate with a kaon or pion. 
The  \Lb invariant mass, $m(\Lc h^-)$, is required to be larger than $5500\mevcc$ to reject background due to partially reconstructed \Lb decays.

Two types of background peaking in the signal mass region are identified.
For the first type, $\Dz$ or $\jpsi$ mesons are observed in the invariant-mass distributions of the two charged companion tracks of $\Lb$ and $\Lc$ decays.
The second type involves a genuine $\KS$ ($\Lz$) decay reconstructed as the $\Lz$ ($\KS$) decay.
These background candidates are suppressed using information from particle identification (PID) detectors or rejected by specific vetoes in the corresponding mass spectra.
A boosted decision tree (BDT) classifier implemented in the TMVA toolkit~\cite{TMVA4} is then used to  separate the \Lb signal from the background of random combinations of final-state particles.
The BDT analysis is performed independently for $\Lc\to p\KS$ and $\Lc\to\Lz h^+$ decays.
Each BDT classifier is trained on simulated signal decays and background from data in the high-mass region $m(\Lc h^-)>5900\mevcc$, 
using a combination of kinematic, topological and isolation variables of the $\Lb$, $\Lc$, $\Lz$ or $\KS$ hadrons.
In the final stage of the event selection,
a simultaneous optimization of   the final-state PID and BDT classifier requirements is performed to maximize the figure of merit, $N_S^{2}/(N_S+N_B)^{3/2}$,
chosen to favour a high signal purity with small decay-parameter uncertainties.
Here, 
$N_S$ and $N_B$ represent the signal and background yields in the signal region chosen to be $\pm32\mevcc$ around the known \Lb mass~\cite{ParticleDataGroup:2024cfk}, estimated with simulated signal decays and data in the high-mass region.
The \Lb invariant mass is calculated with a kinematic fit~\cite{Hulsbergen:2005pu} constraining the masses of all intermediate particles to their known values
and the \Lb momentum to point back to its best-matched PV.

\begin{figure}[tb]
\begin{center}
\includegraphics[width=0.48\columnwidth]{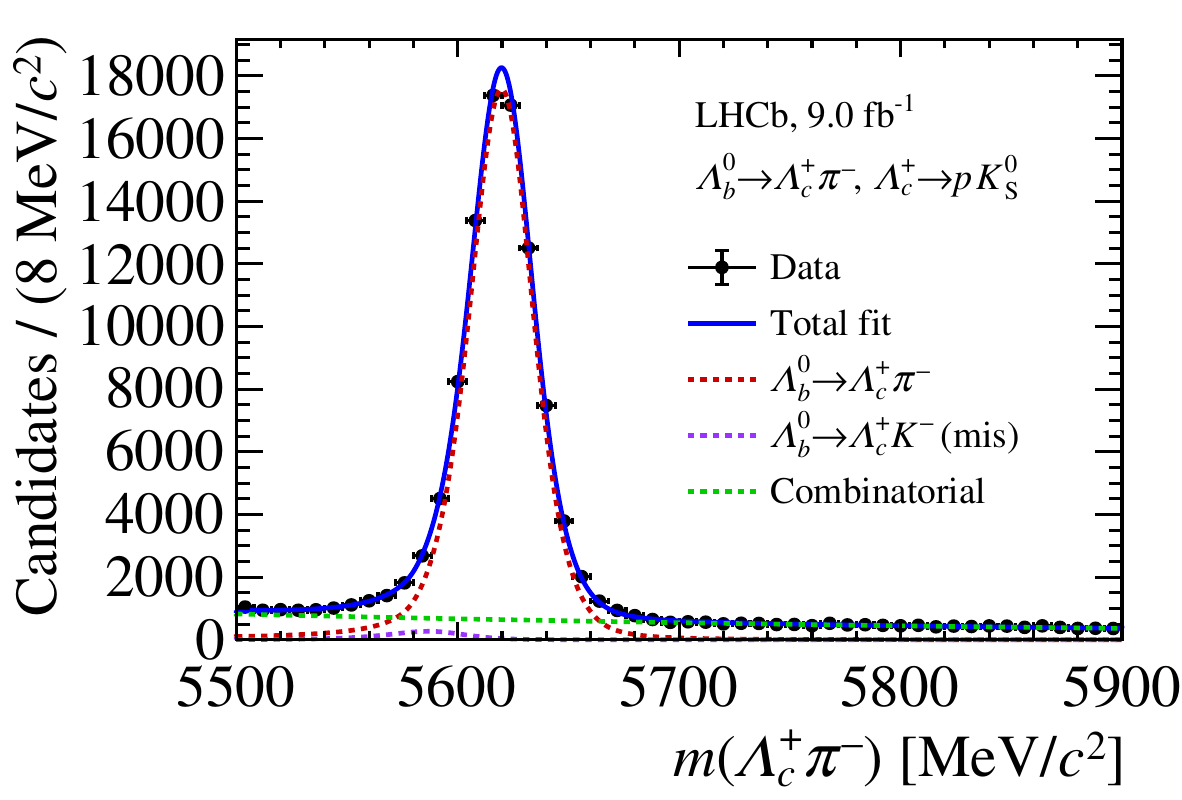}
\includegraphics[width=0.48\columnwidth]{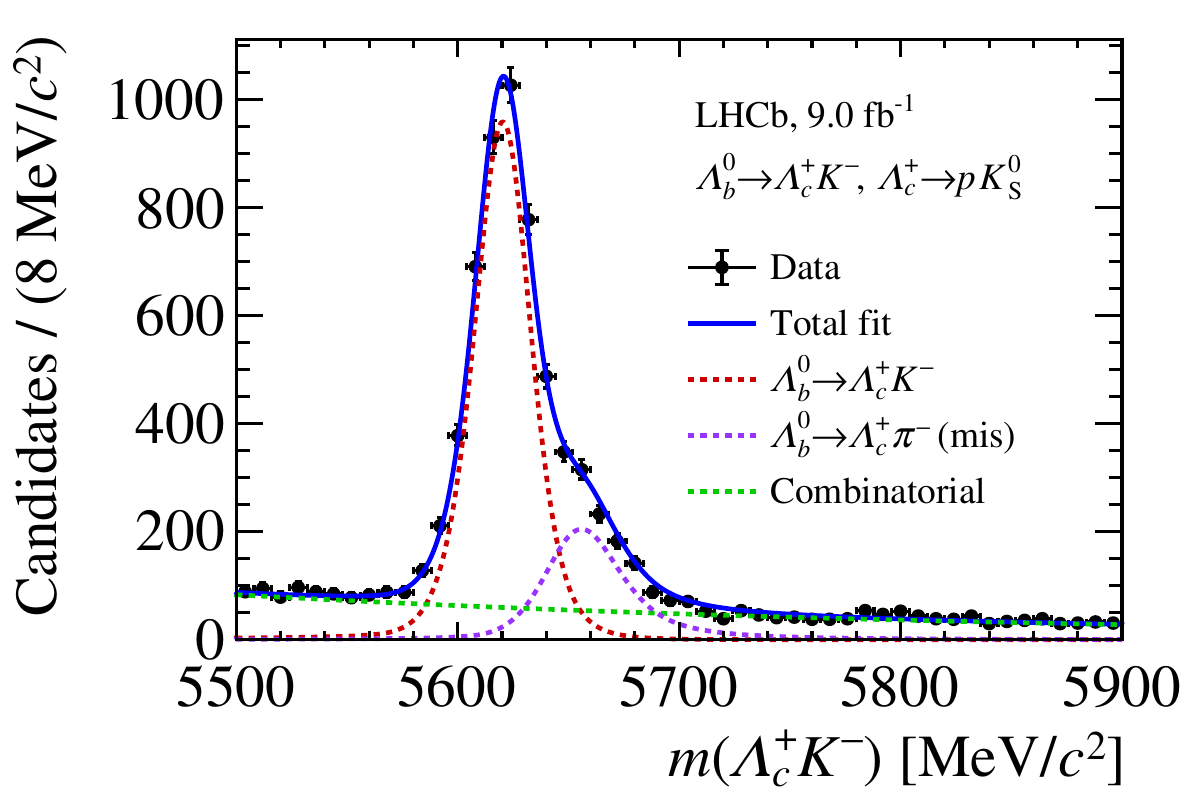}\\
\includegraphics[width=0.48\columnwidth]{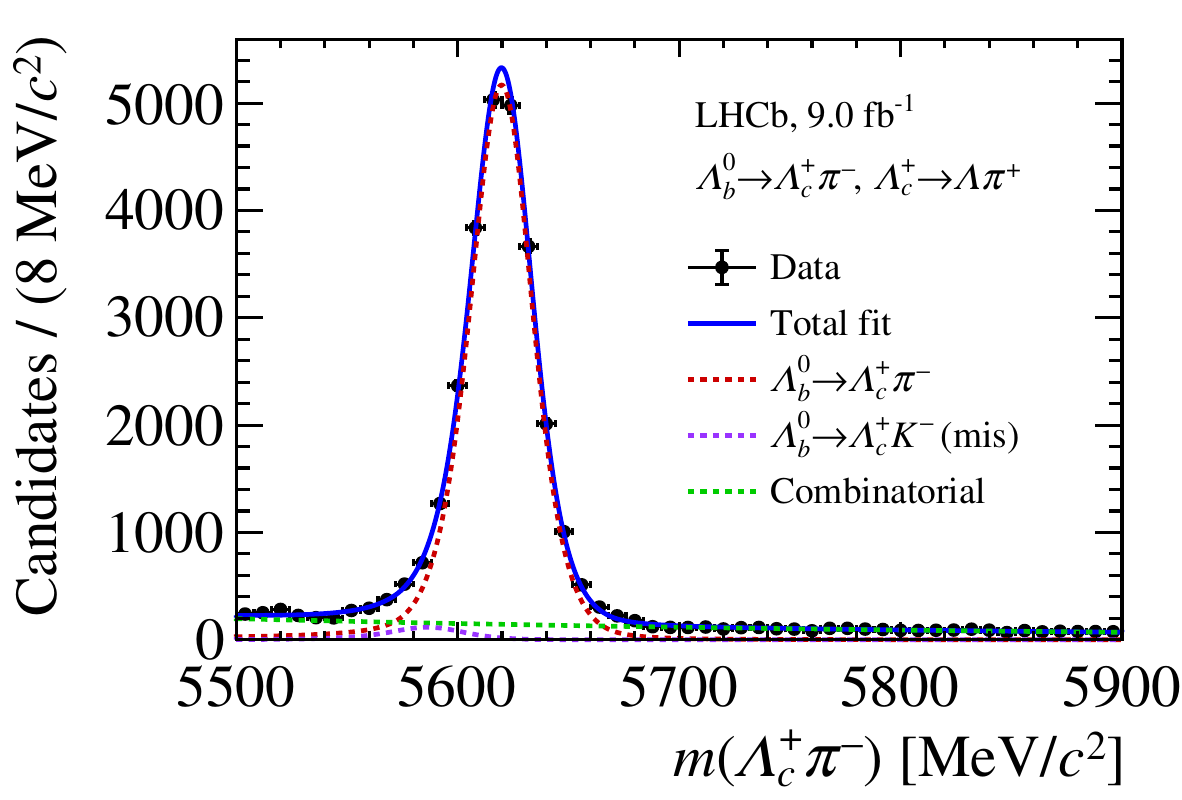}
\includegraphics[width=0.48\columnwidth]{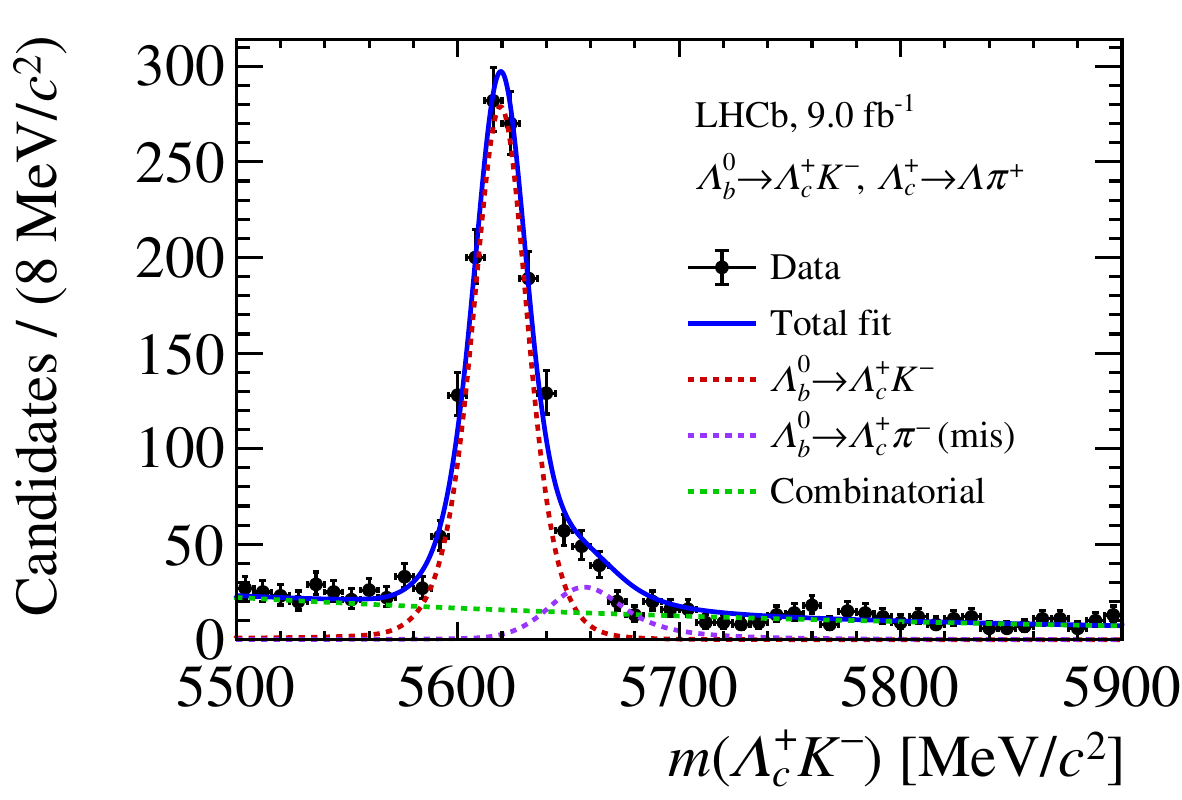}
\includegraphics[width=0.48\columnwidth]{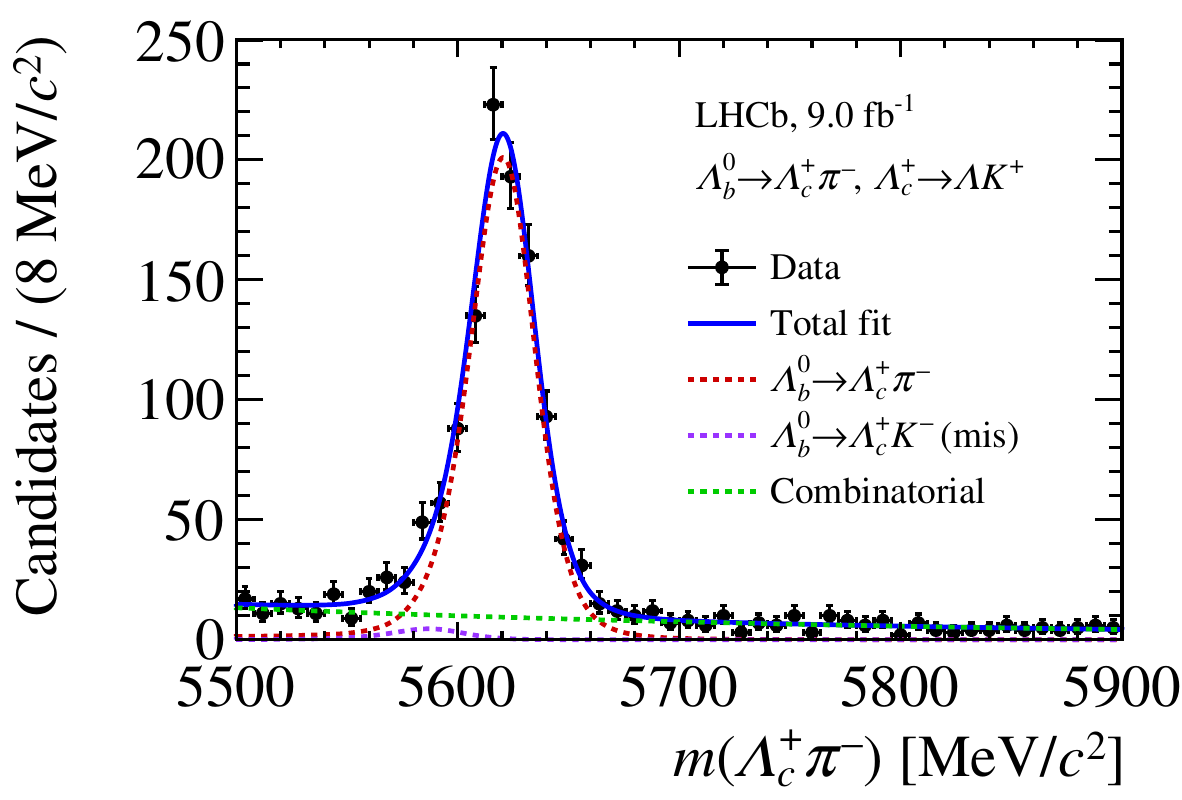}
\end{center}
\caption{The invariant-mass distributions of \Lb candidates reconstructed in the (top left) \LbToLcpiTopKS, (top right) \LbToLcKTopKS, (middle left) \LbToLcpiToLzpi, (middle right) \LbToLcKToLzpi and (bottom)  \LbToLcpiToLzK decays, with the fit results drawn. }
\label{fig:paper_mass}
\end{figure}

The invariant-mass distributions of the five significant $\Lb$ cascade decays to 
$(\proton\KS)\pim$, $(\proton\KS)\Km$, $(\Lz\pip)\pim$, $(\Lz\pip)\Km$ and $(\Lz\Kp)\pim$ final states,
where \Lc decay products are shown in brackets, 
are shown in Fig.~\ref{fig:paper_mass} for candidates passing all selection criteria.
The signal yields of the five decays are determined  to be
$(8.635\pm 0.032)\times10^{4}$, $(4.16\pm0.07)\times10^{3}$, $(2.475\pm0.017)\times10^{4}$, $(1.19\pm0.04)\times10^3$ and $(1.010\pm0.034)\times10^3$, respectively, from 
unbinned maximum-likelihood fits performed to the \Lb mass distributions.
The signal component is described by a  Hypatia function~\cite{MartnezSantos2014} and
the combinatorial background by an exponential function. 
The $\Lb\to\Lc \Km$ decay misidentified as $\Lb\to\Lc \pim$ decay, or vice versa, is also modelled by a Hypatia function,
whose parameters are fixed to those obtained from the simulated samples.
The relative yields of these cross-feed contributions are constrained using relative experimental efficiencies.
For every decay mode, the fit result is used to determine the \sPlot weight for each candidate~\cite{Pivk:2004ty}, applied to subtract the background for the subsequent angular analysis.

\begin{figure}[tb]
\begin{center}
\includegraphics[width=\columnwidth]{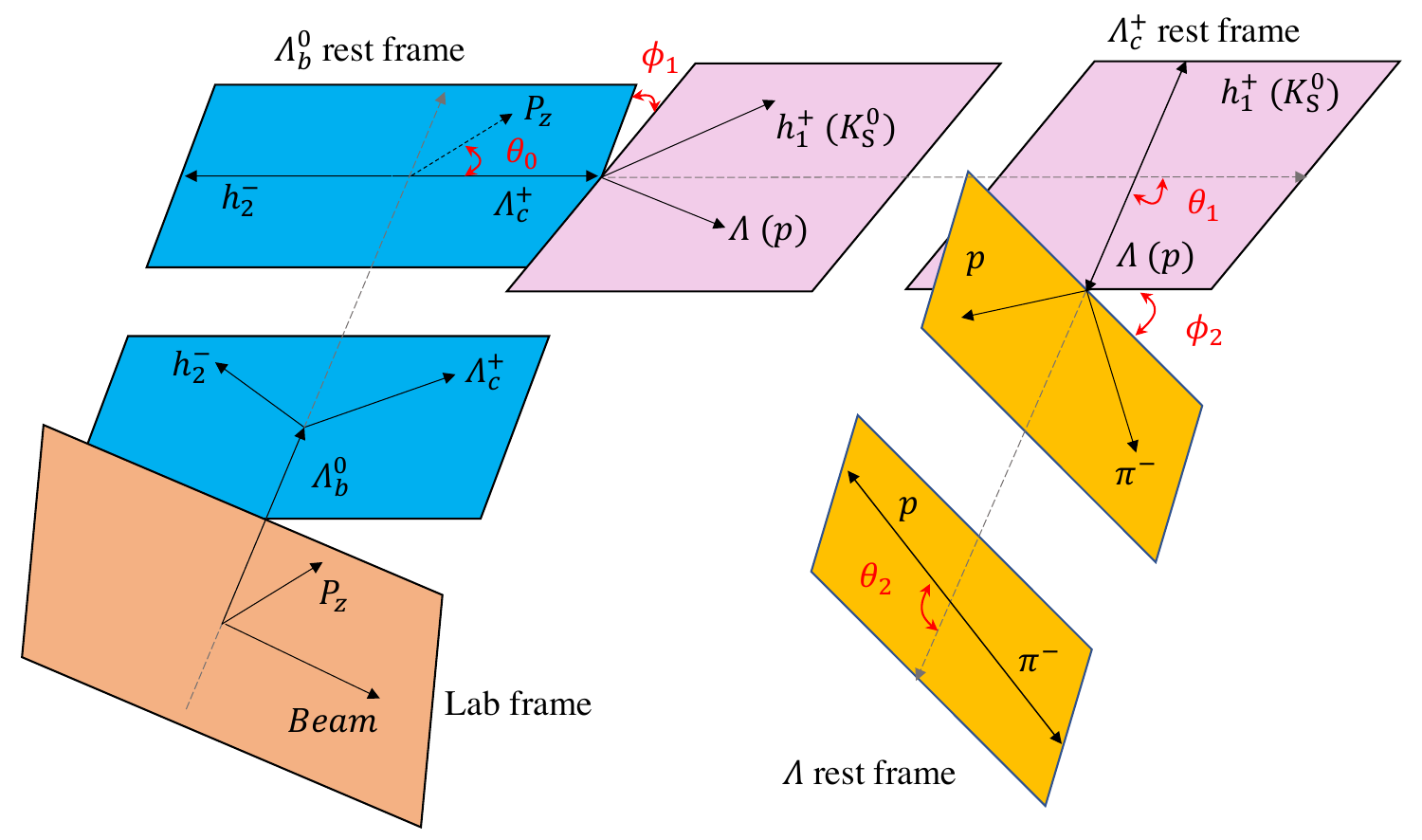}
\end{center}
\caption{Definition of the helicity angles for $\LbnLchLambdahdecay$ and $\LbnLchKspdecay$  decays, where $h_1^+, h_2^-$ denote the kaon or pion.
}
\label{fig:paper_angles}
\end{figure}

The decay parameters are determined by analyzing the angular distributions of the $\Lb$ cascade decays.
The angular variables are calculated with the
\Lb invariant mass constrained to the known value~\cite{ParticleDataGroup:2024cfk}.
The kinematics of the three-step cascade \LbToLchToLh decays are fully described by five angular variables
$\Vec{\Omega}\equiv(\theta_0,\theta_1,\phi_1,\theta_2,\phi_2)$, depicted in Fig.~\ref{fig:paper_angles}.
The variable $\theta_0$ is the polar angle between the normal $\vec{P}_z$ of the production plane formed by the beam and $\Lb$ momenta in the laboratory frame, 
and the $\Lc$ momentum $\vec{p}_\Lc$ in the $\Lb$ rest frame.
The variable $\theta_1$ ($\theta_2$) is the polar angle between $\vec{p}_\Lc$ ($\vec{p}_\proton$) and $\vec{p}_\Lz$, where particle momenta are defined in the rest frames of the $\Lb$ ($\Lz$) and $\Lc$ baryons, respectively.
The variable $\phi_1$ ($\phi_2$) is the angle between the \Lb (\Lz) decay plane and the \Lc decay plane, spanned by the momenta of their respective decay products.
Similarly, for the two-step cascade decays, \LbToLchhTopKS, the kinematics are described by three angular variables
$\Vec{\Omega}\equiv(\theta_0,\theta_1,\phi_1)$, which are the same as the first three variables of the three-step cascade.

The angular distributions can be expanded through the helicity formalism~\cite{Jacob:1959at}.
Based on previous studies at the LHC~\cite{LHCB-PAPER-2012-057,LHCb-PAPER-2020-005,ATLAS:2014swk,CMS:2018wjk},
the $\Lb$ baryon is considered to be unpolarized, 
in which case the angular distributions become uniform in $\theta_0$ and $\phi_1$.
The impact of $\Lb$ polarization is considered as a source of systematic uncertainty.
The reduced angular distributions are thus expressed as
\begin{equation}
\begin{aligned}
	\frac{\mathrm{d}^3\Gamma}{\mathrm{d}\cos\theta_1\mathrm{d}\cos\theta_2\mathrm{d}\phi_2}\propto
	\ &(1+\alphaB\alphaC\cos\theta_1
	+\alphaC\alphaS\cos\theta_2
	+\alphaB\alphaS\cos\theta_1\cos\theta_2\\
	&-\alphaB\gammaC\alphaS\sin\theta_1\sin\theta_2\cos\phi_2
    +\alphaB\betaC\alphaS\sin\theta_1\sin\theta_2\sin\phi_2),
\end{aligned}
\label{equ:paper_3level}
\end{equation}
for \LbToLchToLh~decays,
and
\begin{equation}
	\frac{\mathrm{d}\Gamma}{\mathrm{d}\cos\theta_1}\propto 
	1+\alphaB\alphaCp\cos\theta_1,
\label{equ:paper_2level}
\end{equation}
for \LbToLchhTopKS decays,
where the subscript of the decay parameters denotes the decaying particle.
The decay parameters in this analysis
are determined from simultaneous unbinned maximum-likelihood fits to the five $\Lb$ ($\Lbbar$) cascade decays,
imposing the constraint  $(\alphaCh)^2 + (\betaCh)^2 + (\gammaCh)^2 = 1$. 
The  \betaCh and \gammaCh parameters are related to the \alphaCh and \DeltaCh parameters by 
$\betaCh = \sqrt{1 - (\alphaCh)^2}~\sin{\DeltaCh}$, $\gammaCh = \sqrt{1 - (\alphaCh)^2}~\cos{\DeltaCh}$, 
where \DeltaCh is the phase difference between the two helicity amplitudes of the $\Lc\to\Lz h^+$ decay.
This leads to two equivalent sets of fit parameters for a $\Lc\to\Lz h^+$ decay.
The fit is performed for each set of parameters independently to directly determine their values and uncertainties.
To test \CP violation, an additional joint fit of \Lb and \Lbbar samples is applied with \CP-related fit parameters, which are the \CP asymmetries $A_{\alpha}$, $R_{\beta}$, and \CP averages \mbox{$\langle\alpha\rangle\equiv (\alpha-\bar{\alpha})/2$}, \mbox{$R'_{\beta} \equiv (\beta-\bar{\beta}) / (\alpha-\bar{\alpha})$}.
At leading order, the weak and strong phase differences are determined using $R_{\beta}=\tan\Delta{\phi} $ and $R'_{\beta}=\tan\Delta{\delta}$~\cite{PhysRevD.34.833},
and the quadrant of phases can be determined using Eq.~(45) in Ref.~\cite{Zhong:2024qqs}.

\begin{figure}[tb]
\begin{center}
\includegraphics[width=0.48\columnwidth]{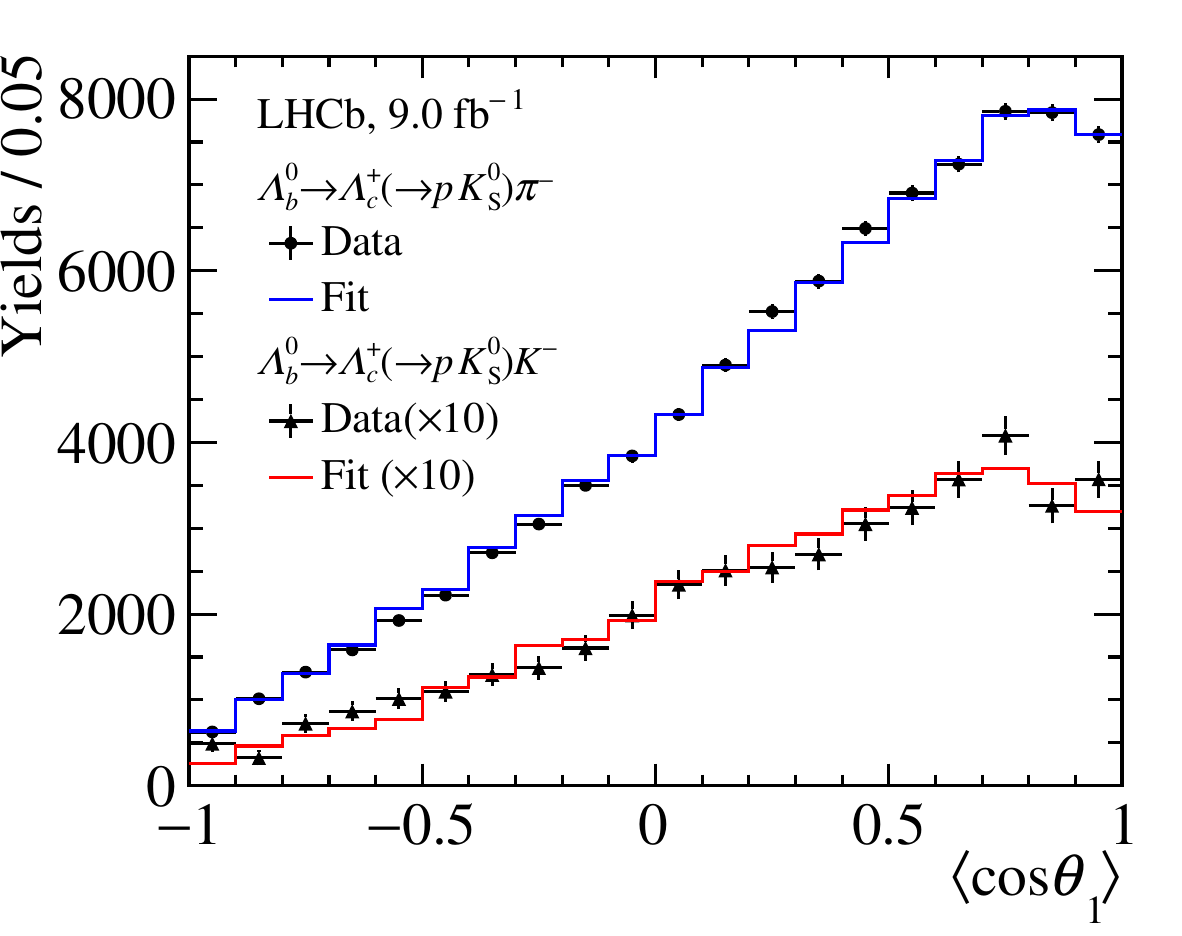}
\includegraphics[width=0.48\columnwidth]{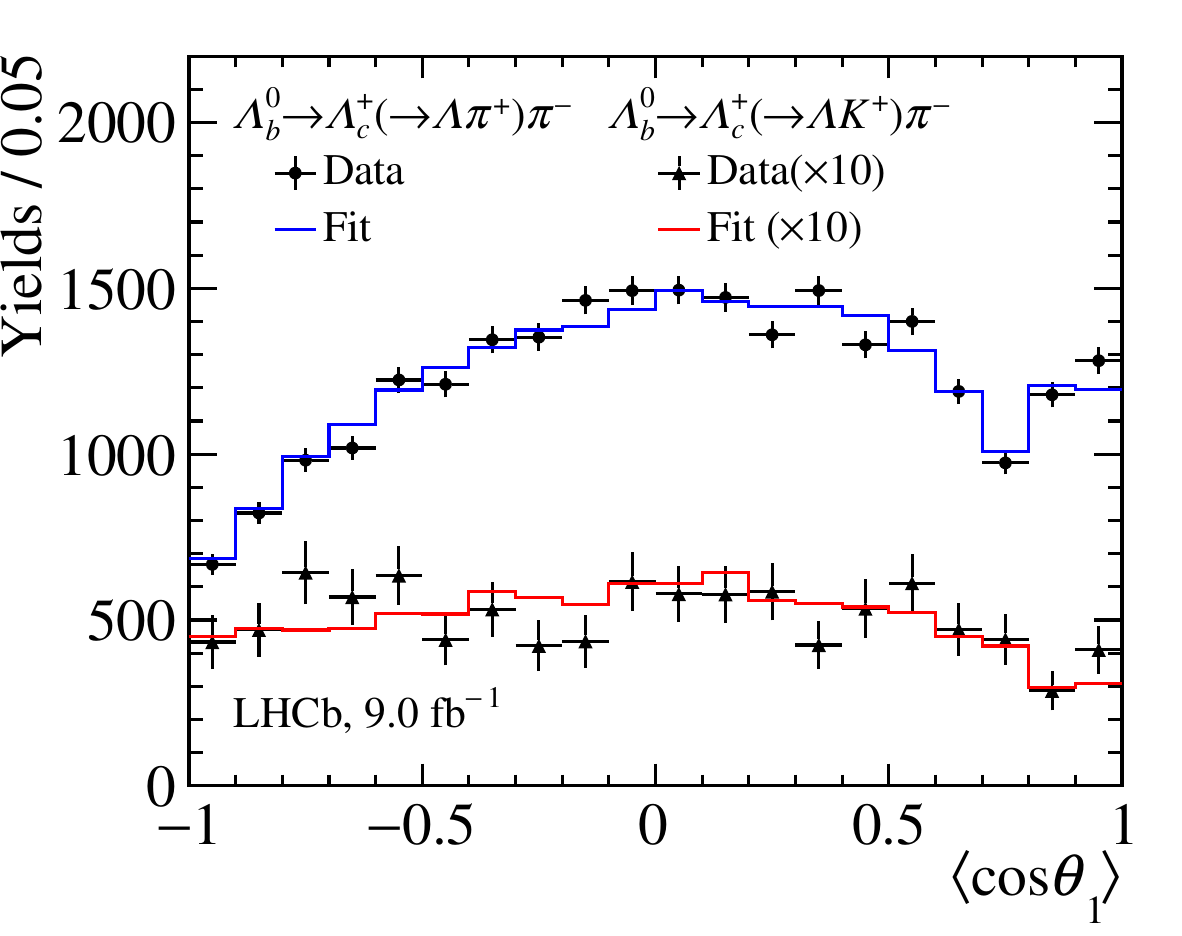}
\includegraphics[width=0.48\columnwidth]{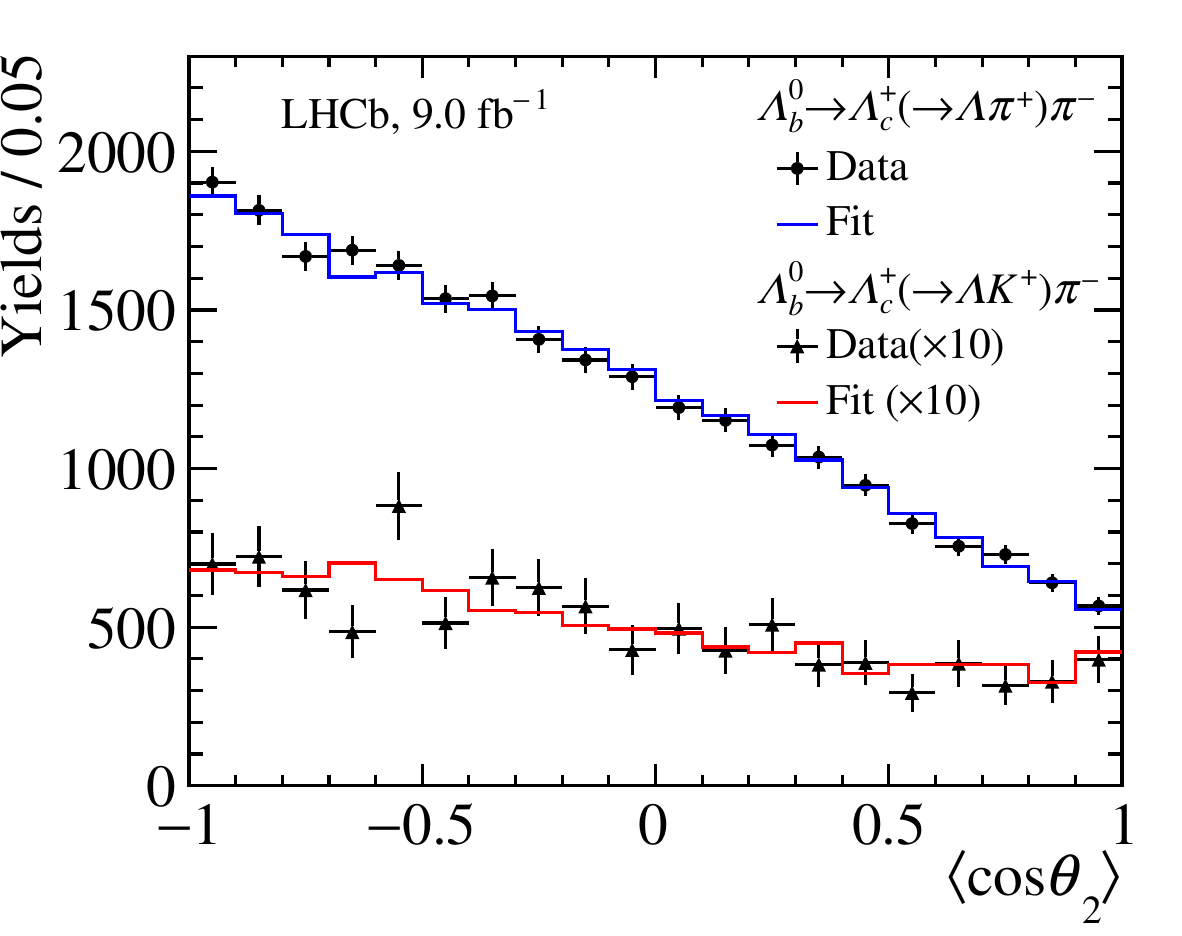}
\includegraphics[width=0.48\columnwidth]{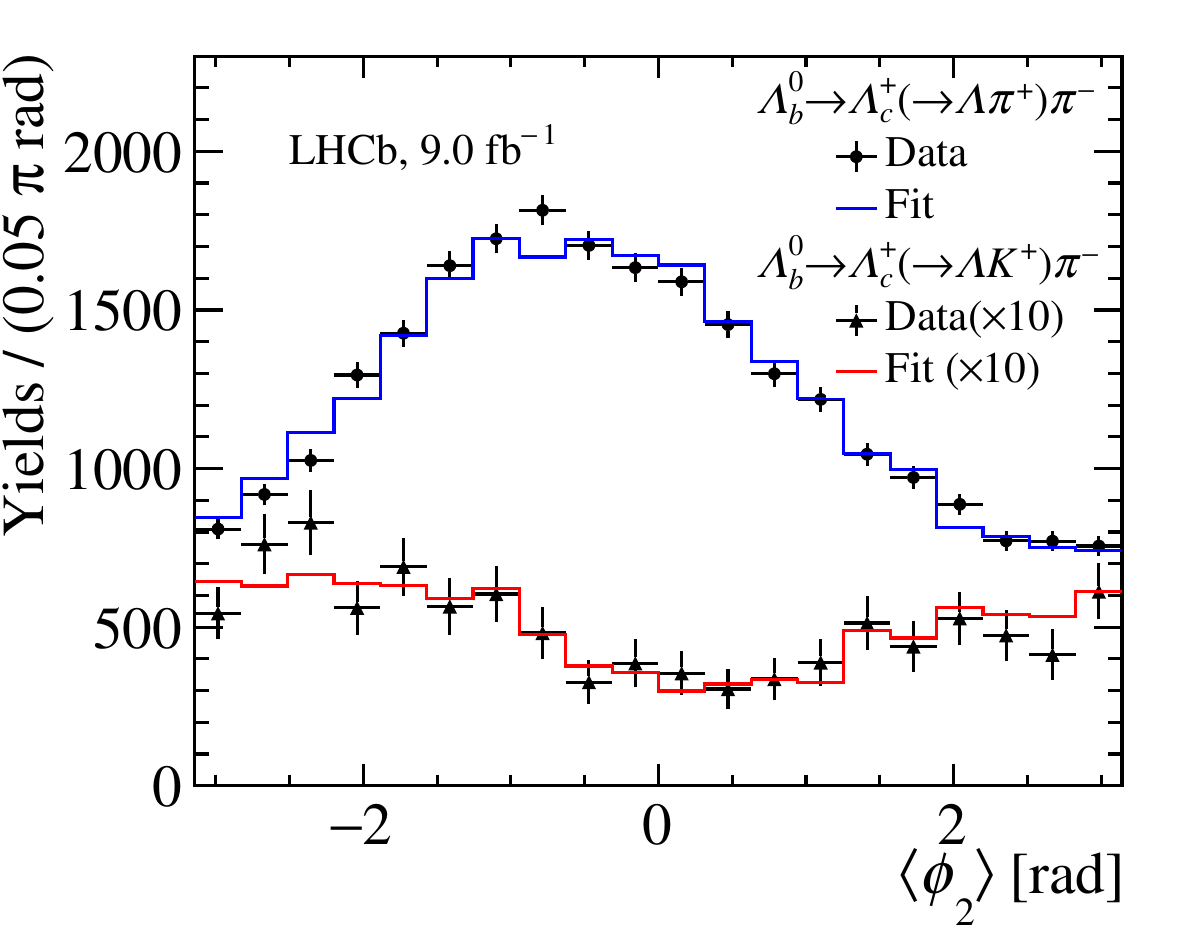}
\end{center}
\caption{Distributions of  (top left) the $\langle\cos\theta_1\rangle$ angle of the \LbToLchTopKS decays,  and the (top right) $\langle\cos\theta_1\rangle$,  (bottom left) $\langle\cos\theta_2\rangle$ and   (bottom right) $\langle\phi_2\rangle$ angles of the \LbToLcpiToLzh decays.
The angular brackets denote that the \Lb and \Lbbar samples are merged, where the $\phi_2$ signs are also flipped for \Lbbar samples.
Points with error bars correspond to background-subtracted data using the \sPlot technique.
}
\label{fig:paper_fit}
\end{figure}

The logarithm of the likelihood function ($\log\mathcal{L}$) is constructed as
\begin{equation}
    \log{\mathcal{L}}(\Vec{\nu}) = \sum_{k=1}^{5} \left(
    \mathcal{C}_k 
    \sum_{i=1}^{N_k} {w}_{k,i}\times\log\left[ \mathcal{P}_k(\Vec{\Omega}_k^i|\Vec{\nu})\right]\right),
\label{equ:paper_ll}
\end{equation}
where $\Vec{\nu}$ is the set of decay parameters, $\Vec{\Omega}$ is the set of angular variables,
and $\mathcal{P}(\Vec{\Omega}|\Vec{\nu})$  represents the signal probability density function (PDF).
The subscript $k$ runs over the five  $\Lb$ cascade decays, 
and the subscript $i$ runs over all the $N_k$ candidates of the $k$-th decay.
The \sPlot weight $w_{k,i}$ in the $\log\mathcal{L}$ is used to remove the contribution of background candidates~\cite{Pivk:2004ty}, while
the constants $\mathcal{C}_k\equiv\sum_{i\in  \text{data}_k}  {w}_{k,i}/\sum_{i\in  \text{data}_k}  {w}_{k,i}^2$ are scale factors needed to correct the obtained statistical uncertainties~\cite{Langenbruch:2019nwe}.
The signal PDF $\mathcal{P}_k(\Vec{\Omega}_k|\Vec{\nu})$  is formulated as
\begin{equation}
    \mathcal{P}_k(\Vec{\Omega}_k|\Vec{\nu}) = \frac{\epsilon_k(\Vec{\Omega}_k) \cdot f_k(\vec{\Omega}_k|\Vec{\nu})}
    {\int \mathrm{d}\Vec{\Omega}_k~\epsilon_k(\Vec{\Omega}_k) \cdot f_k(\vec{\Omega}_k|\Vec{\nu})},
\label{equ:paper_pdf}
\end{equation}
where $f_k(\vec{\Omega}_k|\Vec{\nu})$ represents the angular distribution given in 
Eq.~\ref{equ:paper_3level} or~\ref{equ:paper_2level},
and $\epsilon_k(\Vec{\Omega}_k)$ is the angular acceptance~\cite{LHCb-PAPER-2013-008}. 
The denominator is calculated numerically using the Monte Carlo integration method
beginning with the corresponding simulated signal decays after full selection~\cite{1998AcNum...7....1C, LHCb-PAPER-2013-008}.
The distributions of the $\Lb$ transverse momentum and pseudorapidity, and the number of
tracks per event in the simulation samples are corrected to match those in data.
In Fig.~\ref{fig:paper_fit}, the angular distributions of  $\LbToLchTopKS$ and $\LbToLcpiToLzh$ decays are shown, superimposed by the fit result. 
Distributions for all decays are provided in Ref.~\cite{suppPRL}.
A binned $\chisq$ test between the data and the fit gives a p-value of 28\%.

Various sources of systematic uncertainty on the decay parameters are studied.
Possible biases introduced by the angular fit method are evaluated using pseudoexperiments. 
Mass and angular distributions of pseudosamples, including possible correlations, are generated according to the baseline fit results,
and then the whole fit procedure is repeated to extract decay parameters.
The parameter's systematic uncertainty is taken to be the mean of its pull distribution times its nominal statistical uncertainty.
The \sPlot method is used to subtract the background, 
hence the choice of the invariant-mass fit model introduces systematic uncertainties.
These are estimated by repeating the invariant-mass fit with alternative fit models,
including alternative descriptions of mass-shape functions and removing the constraints on yields,
then using the corresponding updated \sPlot weights to determine decay parameters.
As the PID variables in simulation samples are calibrated to match data~\cite{LHCb-PUB-2016-021,LHCb-DP-2018-001}, 
the uncertainty on the calibration procedure introduces systematic uncertainties which are estimated with alternative calibration configurations. 
The limited size of simulation samples introduces an uncertainty on the efficiency propagated to the decay parameters, 
which is estimated with bootstrapped pseudoexperiments~\cite{efron:1979}. 
The influence of the production asymmetry for $\Lb$ baryons and detection asymmetries on the final-state particles~\cite{LHCb-PAPER-2016-062, LHCb-PAPER-2021-016, LHCb-PAPER-2022-024} are taken into account.
Following the prescription of \CP measurements~\cite{LHCb-PAPER-2018-025, LHCb-PAPER-2018-001},
these asymmetries are introduced in the angular acceptance, 
and the angular fit is repeated to verify their impact on the measurements. 
The polarization of \Lb baryons is considered as a source of systematic uncertainty.
The angular fit is repeated with additional terms in the PDF incorporating the transverse polarization measured by \lhcb~\cite{LHCb-PAPER-2020-005}
(see appendix for details on this PDF).
The impact of the experimental angular resolution is considered as a systematic uncertainty and found to be negligible.
The spin of the \Lz baryon undergoes a precession in the magnetic field of the detector, 
which modifies its angular distribution depending on the decay length~\cite{Botella:2016ksl}.
The systematic uncertainty arising from the precession is examined using pseudoexperiments,
and found to be negligible.
A summary of the contributions from the various sources is given in Ref.~\cite{suppPRL}.
The systematic uncertainties from different sources are added in quadrature, 
resulting in totals that are smaller than the statistical uncertainties.

\begin{table}[tb]
\centering
\caption{Measurements of $\alpha$ parameters and their \CP asymmetries for $\Lb\to\Lc\pim$, $\Lb\to\Lc\Km$, $\Lc\to\Lz\pip$, $\Lc\to\Lz\Kp$, $\Lc\to\proton\KS$ and $\Lz\to\proton\pim$ decays. 
The first uncertainties are statistical and the second are systematic.
}
\resizebox{1 \columnwidth}{!}{
\begin{tabular}{ c c c c c }
\hline \hline
Decay & $\alpha$ & $\bar{\alpha}$ & $\langle\alpha\rangle$ & $A_{\alpha}$ \\ \hline
$\Lb\to\Lc\pim$ & $-1.010\pm0.011\pm0.003$ & \hspace{9pt}$0.996\pm0.011\pm0.003$ & $-1.003\pm0.008\pm0.005$ & \hspace{9pt}$0.007\pm0.008\pm0.005$ \\
$\Lb\to\Lc\Km$  & $-0.933\pm0.042\pm0.014$ & \hspace{9pt}$0.995\pm0.036\pm0.013$ & $-0.964\pm0.028\pm0.015$ & $-0.032\pm0.029\pm0.006$ \\ \hline
$\Lc\to\Lz\pip$ & $-0.782\pm0.009\pm0.004$ & \hspace{9pt}$0.787\pm0.009\pm0.003$ & $-0.785\pm0.006\pm0.003$ &  $-0.003\pm0.008\pm0.002$ \\
$\Lc\to\Lz\Kp$ & $-0.569\pm0.059\pm0.028$ & \hspace{9pt}$0.464\pm0.058\pm0.017$ & $-0.516\pm0.041\pm0.021$ &  \hspace{9pt}$0.102\pm0.080\pm0.023$ \\
$\Lc\to\proton\KS$ & $-0.744\pm0.012\pm0.009$ & \hspace{9pt}$0.765\pm0.012\pm0.007$ & $-0.754\pm0.008\pm0.006$ & $-0.014\pm0.011\pm0.008$ \\ \hline
$\Lz\to\proton\pim$
& \hspace{9pt}$0.717\pm0.017\pm0.009$ & $-0.748\pm0.016\pm0.007$ & \hspace{9pt}$0.733\pm0.012\pm0.006$ & $-0.022\pm0.016\pm0.007$ \\
\hline \hline
\end{tabular}
}
\label{tab:paper_alpha}
\end{table}

\begin{table}[tb]
\centering
\caption{Measurements of the decay parameters $\beta$ and $\gamma$, the phase difference $\Delta$, the \CP asymmetry $R_\beta$ and the \CP average $R'_\beta$ for $ \Lc\to\Lz\pip $, $ \Lc\to\Lz\Kp $ decays and their charge-conjugated decays. 
The first uncertainties are statistical and the second are systematic.
}
\begin{tabular}{ l c c}
\hline \hline
Decay & $\Lc\to\Lz\pip$ & $\Lc\to\Lz\Kp$\\
\hline
$\beta$       & \hspace{9pt}$0.368\pm0.019\pm0.008$ & $\hspace{9pt}0.35\pm0.12\pm0.04$ \\
$\bar{\beta}$ & $-0.387\pm0.018\pm0.010$            & $-0.32\pm0.11\pm0.03$ \\
\hline
$\gamma$      &\hspace{9pt}$0.502\pm0.016\pm0.006$  &$-0.743\pm0.067\pm0.024$\\
$\bar{\gamma}$&\hspace{9pt}$0.480\pm0.016\pm0.007$  &$-0.828\pm0.049\pm0.013$\\
\hline
$\Delta$~(rad)      &\hspace{9pt}$0.633\pm0.036\pm0.013$  &\hspace{9pt}$2.70\pm0.17\pm0.04$\\
$\bar{\Delta}$~(rad)&$-0.678\pm0.035\pm0.013$             &$-2.78\pm0.13\pm0.03$\\
\hline
$R_{\beta}$   &\hspace{9pt}$0.012\pm0.017\pm0.005$   &$-0.04\pm0.15\pm0.02$\\
$R'_{\beta}$  & $-0.481\pm0.019\pm0.009$   &$-0.65\pm0.17\pm0.07$\\
\hline \hline
\end{tabular}
\label{tab:paper_beta}
\end{table}

The results are listed in Table~\ref{tab:paper_alpha} for the $\alpha$ parameters of \Lb, \Lc and \Lz decays, 
and in Table~\ref{tab:paper_beta} for the $\beta$ and $\gamma$ parameters of $\Lc\to\Lz h^+$ decays.
The \CP-related parameters are also obtained,
and no \CP violation is found.
This is the first measurement of the parity-violating parameters of two-body \Lb decays into a spin-half baryon and a pseudoscalar meson.
The results of the \alphaB decay parameters  are close to $-1$, 
suggesting that \Lc baryons in $\Lb\to\Lc h^-$ decays are almost fully longitudinally polarized, which corresponds to the $V-A$ nature of weak decays and supports the factorization hypothesis in theoretical calculations~\cite{PhysRevD.56.2799}.
The $\Lc$ decay parameters are consistent with, and more precise than, the \belle~\cite{Belle:2022uod} and \besiii~\cite{BESIII:2019odb} results.
The $\alphaC$ parameters are found to significantly deviate from $-1$, which may suggest that nonfactorizable contributions are substantial in hadronic decays of charm baryons.
The $\beta, \gamma$ and $\Delta$ parameters of $\Lc\to\Lz h^+$ decays are precisely measured for the first time, 
and will serve as essential inputs to theoretical models~\cite{Zhong:2024qqs}.
The weak and strong phase differences are determined to be
$\Delta \phi=0.01\pm 0.02\rad$  and $\Delta\delta =2.693\pm 0.017\rad$  for the $\Lc\to\Lz\pip$ decay,
and $\Delta \phi=-0.03\pm 0.15\rad$  and $\Delta\delta =2.57\pm 0.19\rad$ for the $\Lc\to\Lz\Kp$ decay.
The $\alpha$ parameter and the corresponding \CP asymmetry of the $\Lz\to\proton\pim$ decay in this analysis are consistent with the \besiii results~\cite{BESIII:2018cnd,PhysRevLett.129.131801}.

In conclusion, based on $pp$ collision data collected by the~\lhcb experiment, corresponding to an integrated luminosity of $9\invfb$,
a comprehensive study of the angular distributions in \Lb cascade decays
is performed.
The analysis provides the first measurements of the decay parameters for $\Lb\to\Lc h^-$ decays, and the most precise measurements for the $\Lc$ decay parameters.  
The weak and strong phase differences for $\Lc\to\Lz h^+$ decays are also determined.
The \CP asymmetries are studied between the decay parameters of baryon and antibaryon decays, and no hint of \CP violation is observed.
The results provide valuable insights into the weak decay dynamics of baryons.

\section*{Acknowledgements}
%
%
\noindent We express our gratitude to our colleagues in the CERN
accelerator departments for the excellent performance of the LHC. We
thank the technical and administrative staff at the LHCb
institutes.
We acknowledge support from CERN and from the national agencies:
CAPES, CNPq, FAPERJ and FINEP (Brazil); 
MOST and NSFC (China); 
CNRS/IN2P3 (France); 
BMBF, DFG and MPG (Germany); 
INFN (Italy); 
NWO (Netherlands); 
MNiSW and NCN (Poland); 
MCID/IFA (Romania); 
MICIU and AEI (Spain);
SNSF and SER (Switzerland); 
NASU (Ukraine); 
STFC (United Kingdom); 
DOE NP and NSF (USA).
We acknowledge the computing resources that are provided by CERN, IN2P3
(France), KIT and DESY (Germany), INFN (Italy), SURF (Netherlands),
PIC (Spain), GridPP (United Kingdom), 
CSCS (Switzerland), IFIN-HH (Romania), CBPF (Brazil),
and Polish WLCG (Poland).
We are indebted to the communities behind the multiple open-source
software packages on which we depend.
Individual groups or members have received support from
ARC and ARDC (Australia);
Key Research Program of Frontier Sciences of CAS, CAS PIFI, CAS CCEPP, 
Fundamental Research Funds for the Central Universities, 
and Sci. \& Tech. Program of Guangzhou (China);
Minciencias (Colombia);
EPLANET, Marie Sk\l{}odowska-Curie Actions, ERC and NextGenerationEU (European Union);
A*MIDEX, ANR, IPhU and Labex P2IO, and R\'{e}gion Auvergne-Rh\^{o}ne-Alpes (France);
AvH Foundation (Germany);
ICSC (Italy); 
Severo Ochoa and Mar\'ia de Maeztu Units of Excellence, GVA, XuntaGal, GENCAT, InTalent-Inditex and Prog. ~Atracci\'on Talento CM (Spain);
SRC (Sweden);
the Leverhulme Trust, the Royal Society
 and UKRI (United Kingdom).

\setcounter{equation}{0}

\clearpage
\section*{End matter}
\section{Angular distributions}

The helicity formalism is employed to describe the angular distributions of the decays in this Letter.
For the decay of a spin-half baryon to a spin-half baryon and a pseudoscalar
meson, two helicity amplitudes are involved with the respective couplings $H_\pm$, 
where the subscript represents the sign of the helicity of the final-state spin-half baryon.
The helicity couplings are related to the S-wave ($s$) and P-wave ($p$) couplings as $s=(H_+ + H_-)/\sqrt{2}$ and $p=(H_+-H_-)/\sqrt{2}$.
The decay parameters are defined using the helicity amplitudes as
\begin{equation}
    \alpha = \frac{|H_+|^2 - |H_-|^2}{|H_+|^2 + |H_-|^2},\quad \beta = \sqrt{1 - \alpha^2} \sin \Delta,\quad
    \gamma = \sqrt{1 - \alpha^2} \cos \Delta,
\label{eqn:paper_helicty}
\end{equation}
where $\Delta = \arg(H_+ / H_-)$ is the phase angle difference between the two helicity amplitudes.

The angular distribution  is determined by the sum of all possible helicity  amplitudes as 
\begin{equation}
    \frac{\mathrm{d}\Gamma}{\mathrm{d}\Omega} \propto |M|^2 = \sum_{\lambda_0, \lambda'_0, \lambda_n} \rho_{\lambda_0, \lambda'_0}~M_{\lambda_0, \lambda_n}~M^*_{\lambda'_0, \lambda_n},
\label{eqn:paper_amplitude_sum}
\end{equation}
where $\lambda_0^{(')}$ and $ \lambda_n$ run over the helicities of the initial and final baryons, $\rho_{\lambda_0, \lambda'_0}$ is the polarization density matrix of the decaying baryon,
and $M_{\lambda_0, \lambda_n}$, $M^*_{\lambda'_0, \lambda_n}$ are the amplitude matrix elements.

For the \Lb baryon promptly produced  in $pp$ collisions, 
the possible polarization is expected to be perpendicular to the production plane due to parity conservation in strong interactions. 
Defining the polarization axis as the $z$-axis, and the magnitude of the polarization as $P_z$, 
the polarization  density matrix is expressed as
\begin{equation}
    \rho = \begin{pmatrix}
            1 + P_z &  0\\
            0 & 1-P_z 
            \end{pmatrix}.
\label{eqn:paper_density}
\end{equation}

\subsection[Angular distribution for LbToLchTopKS decays]{Angular distribution for $\boldsymbol{\LbToLchTopKS}$ decays}

For \LbToLchTopKS decays,
the helicity amplitude is determined as
\begin{equation}
\label{equ:paper_2levelderivation}
    M_{\lambda_b, \lambda_\proton} = \sum_{\lambda_c} H^{b}_{\lambda_c}~d^\frac{1}{2}_{\lambda_b, \lambda_c}(\theta_0) \cdot
    H^{c}_{\lambda_\proton}~e^{i\lambda_c\phi_1}~d^\frac{1}{2}_{\lambda_c, \lambda_\proton}(\theta_1),
\end{equation}
where $d^{J}_{\lambda, \lambda'}(\theta)$ is the Wigner d-matrix,
$\lambda_b, \lambda_c$ and $\lambda_\proton$ refer to the helicities of \Lb, \Lc and \proton baryons, and
$H^{b}_{\lambda_c}$ and $H^{c}_{\lambda_\proton}$ are the helicity couplings of \Lb and \Lc decays.
The total amplitude squared is calculated by 
\begin{equation}
\label{equ:paper_2levelderivation2}
    |M|^2 \propto \sum_{\lambda_\proton}\left[ (1 + P_z) \cdot |M_{1/2, \lambda_\proton}|^2 + (1 - P_z) \cdot |M_{-1/2, \lambda_\proton}|^2\right],
\end{equation}
which leads to
\begin{equation}
\label{equ:paper_2levelwithp}
\begin{aligned}
	\frac{\mathrm{d}^3\Gamma}{\mathrm{d}\cos\theta_0\mathrm{d}\cos\theta_1\mathrm{d}\phi_1}\propto
	\ \ \ 
    &1+\alphaB\alphaC\cos\theta_1\\
	&+P_z\cdot (\alphaB\cos\theta_0+\alphaC\cos\theta_0\cos\theta_1\\
	&-\gammaB\alphaC\sin\theta_0\sin\theta_1\cos\phi_1\\
	&+\betaB\alphaC\sin\theta_0\sin\theta_1\sin\phi_1),
\end{aligned}
\end{equation}
where $\alphaB,\betaB,\gammaB$ are the $\Lb$ decay parameters defined by $H_{\pm}^b$, and
$\alphaC$ is the $\Lc$ decay parameter related to $H_{\pm}^c$.

\subsection[Angular distribution for LbToLchToLh decays]{Angular distribution for  $\boldsymbol{\LbToLchToLh}$ decays}

For \LbToLchToLh decays,
the relevant  angles are $(\theta_0, \theta_1, \phi_1, \theta_2, \phi_2)$, which are defined in Fig.~\ref{fig:paper_angles}.
The helicity amplitude is expressed as
\begin{equation}
\label{equ:paper_3levelderivation}
    M_{\lambda_b, \lambda_\proton} = \sum_{\lambda_c} H^{b}_{\lambda_c}~d^\frac{1}{2}_{\lambda_b, \lambda_c}(\theta_0) \cdot
    H^{c}_{\lambda_s}~e^{i\lambda_c\phi_1}~d^\frac{1}{2}_{\lambda_c, \lambda_s}(\theta_1) \cdot
    H^{s}_{\lambda_\proton}~e^{i\lambda_s\phi_2}~d^\frac{1}{2}_{\lambda_s, \lambda_\proton}(\theta_2),
\end{equation}
where $\lambda_s$ refers to the helicity of \Lz baryons, and 
$H^{c}_{\lambda_s}$ and $H^{s}_{\lambda_\proton}$ are the helicity couplings of \Lc and \Lz decays.
The total amplitude is calculated by Eq.~\ref{equ:paper_2levelderivation2}, which leads to
\begin{equation}
\label{equ:paper_3levelwithp}
\begin{aligned}
	&\frac{\mathrm{d}^5\Gamma}{\mathrm{d}\cos\theta_0\mathrm{d}\cos\theta_1\mathrm{d}\phi_1\mathrm{d}\cos\theta_2\mathrm{d}\phi_2}\\
    \propto \quad
	&(1+\alphaB\alphaC\cos\theta_1
	+\alphaC\alphaS\cos\theta_2
	+\alphaB\alphaS\cos\theta_1\cos\theta_2\\
	&-\alphaB\gammaC\alphaS\sin\theta_1\sin\theta_2\cos\phi_2
    +\alphaB\betaC\alphaS\sin\theta_1\sin\theta_2\sin\phi_2)\\
	+P_z&\cdot (\alphaB\cos\theta_0
	+\alphaC\cos\theta_0\cos\theta_1
	+\alphaB\alphaC\alphaS\cos\theta_0\cos\theta_2\\
	&+\alphaS\cos\theta_0\cos\theta_1\cos\theta_2
	-\gammaB\alphaC\sin\theta_0\sin\theta_1\cos\phi_1
	+\betaB\alphaC\sin\theta_0\sin\theta_1\sin\phi_1\\
	&-\gammaC\alphaS\cos\theta_0\sin\theta_1\sin\theta_2\cos\phi_2
	+\betaC\alphaS\cos\theta_0\sin\theta_1\sin\theta_2\sin\phi_2\\
	&-\gammaB\alphaS\sin\theta_0\sin\theta_1\cos\theta_2\cos\phi_1
	+\betaB\alphaS\sin\theta_0\sin\theta_1\cos\theta_2\sin\phi_1\\
    &+\betaB\betaC\alphaS\sin\theta_0\sin\theta_2\cos\phi_1\cos\phi_2
    +\betaB\gammaC\alphaS\sin\theta_0\sin\theta_2\cos\phi_1\sin\phi_2\\
	&+\gammaB\betaC\alphaS\sin\theta_0\sin\theta_2\sin\phi_1\cos\phi_2
    +\gammaB\gammaC\alphaS\sin\theta_0\sin\theta_2\sin\phi_1\sin\phi_2\\
	&-\gammaB\gammaC\alphaS\sin\theta_0\cos\theta_1\sin\theta_2\cos\phi_1\cos\phi_2\\
	&+\gammaB\betaC\alphaS\sin\theta_0\cos\theta_1\sin\theta_2\cos\phi_1\sin\phi_2\\
	&+\betaB\gammaC\alphaS\sin\theta_0\cos\theta_1\sin\theta_2\sin\phi_1\cos\phi_2\\
	&-\betaB\betaC\alphaS\sin\theta_0\cos\theta_1\sin\theta_2\sin\phi_1\sin\phi_2),
\end{aligned}
\end{equation}
where 
$\alphaS$ is the $\Lz$ decay parameter related to $H_{\pm}^s$.
\clearpage



\addcontentsline{toc}{section}{References}
\bibliographystyle{LHCb}
\bibliography{main,standard,LHCb-PAPER,LHCb-CONF,LHCb-DP,LHCb-TDR}

\clearpage
\setcounter{figure}{0}
\setcounter{table}{0}
\setcounter{equation}{0}
\setcounter{page}{1}
\setcounter{section}{0}
\renewcommand\thefigure{S\arabic{figure}}
\renewcommand\thetable{S\arabic{table}} 
{\LARGE \textbf{Supplemental material}}
\section{Summary of systematic uncertainties}
The various sources of systematic uncertainties on the decay parameter measurements, including fit procedure, mass fit model, PID calibration, limited size of simulation samples, production and detection asymmetries and \Lb polarization, are summarized in Table~\ref{tab:paper_sys}, Table~\ref{tab:paper_sys2} and Table~\ref{tab:paper_sys3}.
The total systematic uncertainties correspond to the sum in quadrature of all sources.

\begin{table}[htb]
\centering
\caption{Systematic uncertainties ($\times 10^{-3}$) on $\alpha$ parameters of different decays. For each cell, the four values are for  baryon decays ($\alpha$), antibaryon decays ($\bar{\alpha}$), their averages ($\langle\alpha\rangle$) and  asymmetries ($A_\alpha$).
}
\resizebox{1 \columnwidth}{!}{
\begin{tabular}{ccccccc}
\hline\hline
Sources  & $\Lb\to\Lc\pim$ & $\Lb\to\Lc\Km$ & $\Lc\to\Lz\pip$ & $\Lc\to\Lz\Kp$ & $\Lc\to\proton\KS$ & $\Lz\to\proton\pim$ \\ \hline
Fit & 0.1/0.7/0.5/0.3 & 2.4/6.6/1.5/4.6 & 0.9/0.6/0.1/0.8 & 3.5/0.9/1.9/2.0 & 0.1/0.3/0.0/0.2 & 0.2/0.3/0.1/0.1 \\
Mass & 1.1/0.9/0.9/0.3 & 4.0/6.0/6.2/1.3 & 0.5/0.6/0.5/0.1 & 23.3/6.7/13.8/22.7 & 1.7/1.6/1.7/0.3 & 1.4/1.7/1.2/0.3 \\
PID & 0.4/0.3/0.3/0.0 & 3.6/3.0/3.6/0.6 & 0.5/0.5/0.1/0.1 & 4.2/3.9/4.7/0.9 & 0.8/0.8/0.5/0.1 & 0.9/0.8/0.7/0.1 \\
MC & 2.6/2.3/2.5/0.2 & 12.0/9.2/10.6/1.9 & 2.7/2.6/2.6/0.2 & 14.4/15.2/14.7/4.9 & 4.4/4.3/4.4/0.2 & 5.5/5.1/5.3/0.6 \\
Asym. & 1.0/1.0/0.2/1.0 & 0.7/1.0/0.2/0.9 & 1.2/1.0/0.0/1.4 & 1.2/1.9/0.6/0.3 & 5.0/5.4/0.4/6.9 & 4.0/4.0/0.1/5.5 \\
Polar. & 0.8/0.6/4.4/4.5 & 4.3/0.8/7.6/2.7 & 1.9/1.0/1.2/1.5 & 4.7/1.4/1.4/1.1 & 4.6/0.9/3.0/4.4 & 4.6/0.9/2.8/3.5 \\ \hline
Total & 3.1/2.9/5.2/4.7 & 14.1/13.3/15.0/5.9 & 3.7/2.9/2.9/2.2 & 28.4/17.3/20.8/23.4 & 8.7/7.2/5.6/8.2 & 8.7/6.8/6.2/6.6 \\
\hline \hline
\end{tabular}
}
\label{tab:paper_sys}
\end{table}

\begin{table}[htb]
\centering
\caption{Systematic uncertainties ($\times 10^{-3}$) on the decay parameters $\beta$ and $\gamma$, the phase difference~$\Delta$,  the \CP asymmetry $R_\beta$ and the \CP average $R'_\beta$ for $ \Lc\to\Lz\pip $ decay.
}
\renewcommand{\arraystretch}{1.1}
\begin{tabular}{cccc}
\hline\hline
Sources   & $\beta/\bar{\beta}/R_{\beta}/R'_{\beta}$ & $\gamma/\bar{\gamma}$ & $\Delta/\bar{\Delta}$ (rad)   \\ \hline
Fit & 0.2/0.1/0.2/0.1& 0.6/1.1 & 0.1/2.0\\
Mass & 1.3/1.3/0.5/0.8& 0.5/1.0 & 1.1/1.1\\
PID & 1.1/1.1/0.1/0.9& 0.8/0.7 & 1.2/1.2\\
MC & 6.7/8.8/0.5/8.5& 5.4/6.6 & 11.9/11.0 \\
Asym. & 3.5/4.4/5.2/0.5& 1.1/1.7 & 5.8/7.1\\
Polar. & 0.2/0.3/0.9/1.4& 0.7/1.1 & 1.1/0.2\\ \hline
Total & 7.7/10.0/5.3/8.7 & 5.7/7.1 & 13.4/13.4 \\
\hline \hline
\end{tabular}
\label{tab:paper_sys2}
\end{table}

\begin{table}[htb]
\centering
\caption{Systematic uncertainties ($\times 10^{-3}$) on the decay parameters $\beta$ and $\gamma$, the phase difference~$\Delta$, the \CP asymmetry $R_\beta$ and the \CP average $R'_\beta$ for $ \Lc\to\Lz\Kp $ decay. 
}
\renewcommand{\arraystretch}{1.1}
\begin{tabular}{cccc}
\hline\hline
Sources   & $\beta/\bar{\beta}/R_{\beta}/R'_{\beta}$ & $\gamma/\bar{\gamma}$ & $\Delta/\bar{\Delta}$ (rad) \\ \hline
Fit & 12.0/2.6/9.5/13.2& 8.0/1.4 & 7.0/1.3 \\
Mass  & 12.2/6.2/13.1/25.4& 15.9/3.3 & 9.7/5.9 \\
PID  & 6.9/4.2/2.7/15.2& 1.6/1.6 & 11.1/7.7 \\
MC  & 32.5/26.4/9.2/54.9& 16.0/12.7 & 37.9/27.7 \\
Asym. & 1.4/1.8/3.1/1.1& 1.4/0.4 & 1.8/2.9 \\
Polar. & 0.4/2.4/0.1/0.2& 0.6/1.1 & 3.2/4.7 \\ \hline
Total & 37.4/27.8/19.1/64.1 & 24.0/13.4 & 41.5/29.9 \\
\hline \hline
\end{tabular}
\label{tab:paper_sys3}
\end{table}

\clearpage
\section{Projections of angular distributions}

\begin{figure}[htb]
\begin{center}
\includegraphics[width=0.45\textwidth]{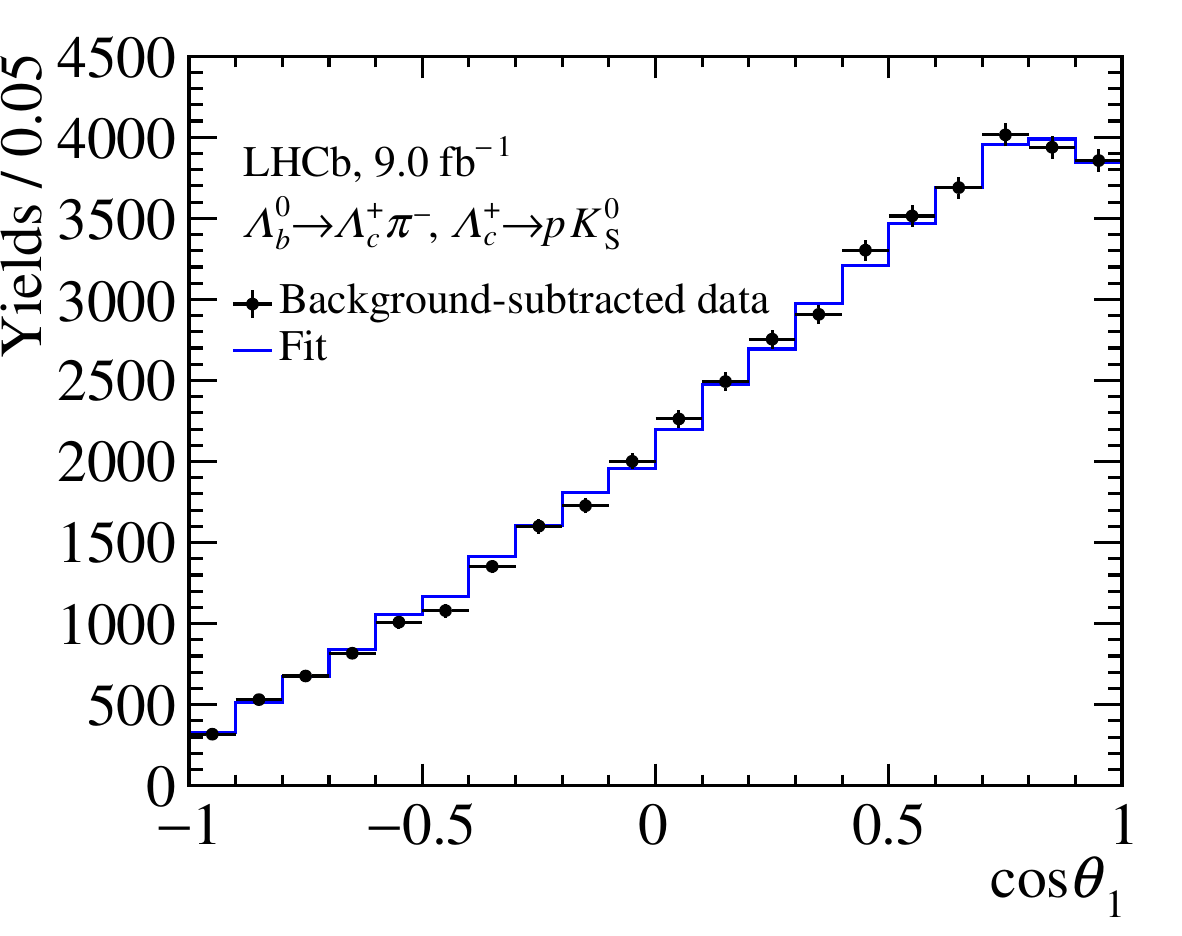}
\includegraphics[width=0.45\textwidth]{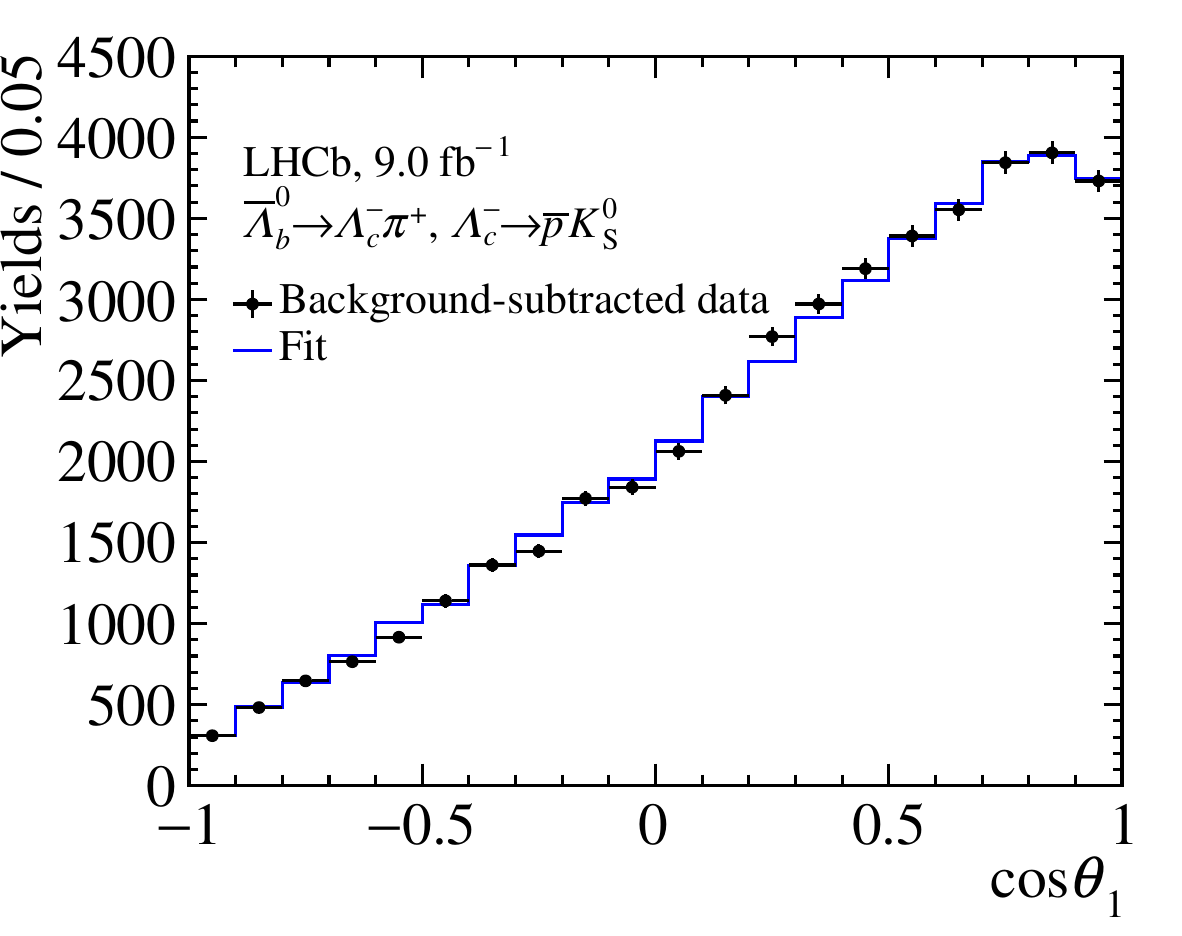}
\end{center}
\caption{Distributions of $\cos\theta_1$  for (left) the \LbToLcpiTopKS decay and (right) its charge-conjugate decay.
}
\label{fig:paper_Lcpi_pKS}
\end{figure}

\begin{figure}[htb]
\begin{center}
\includegraphics[width=0.45\textwidth]{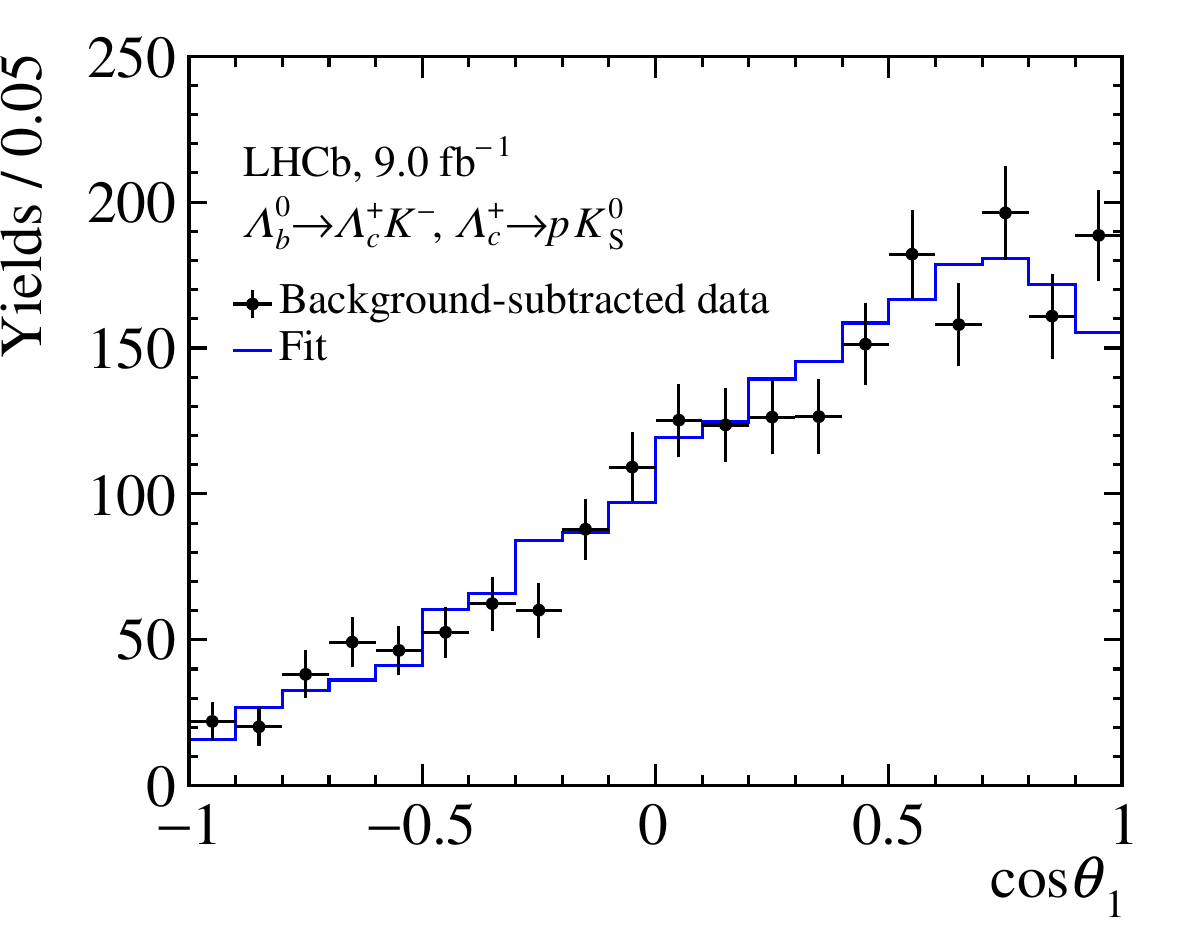}
\includegraphics[width=0.45\textwidth]{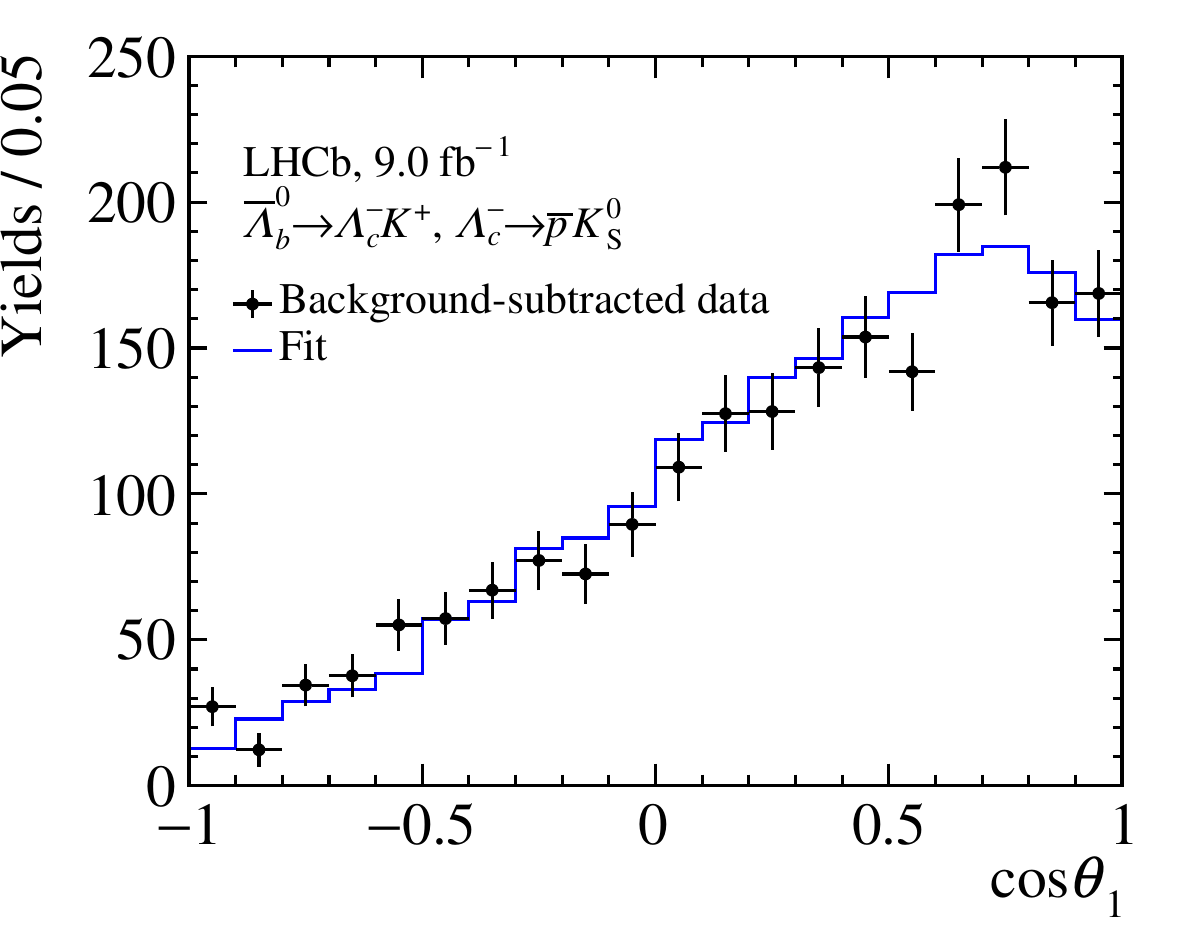}
\end{center}
\caption{Distributions of $\cos\theta_1$  for (left) the \LbToLcKTopKS decay and (right) its charge-conjugate decay.
}
\label{fig:paper_LcK_pKS}
\end{figure}

\begin{figure}[htb]
\begin{center}
\includegraphics[width=0.45\textwidth]{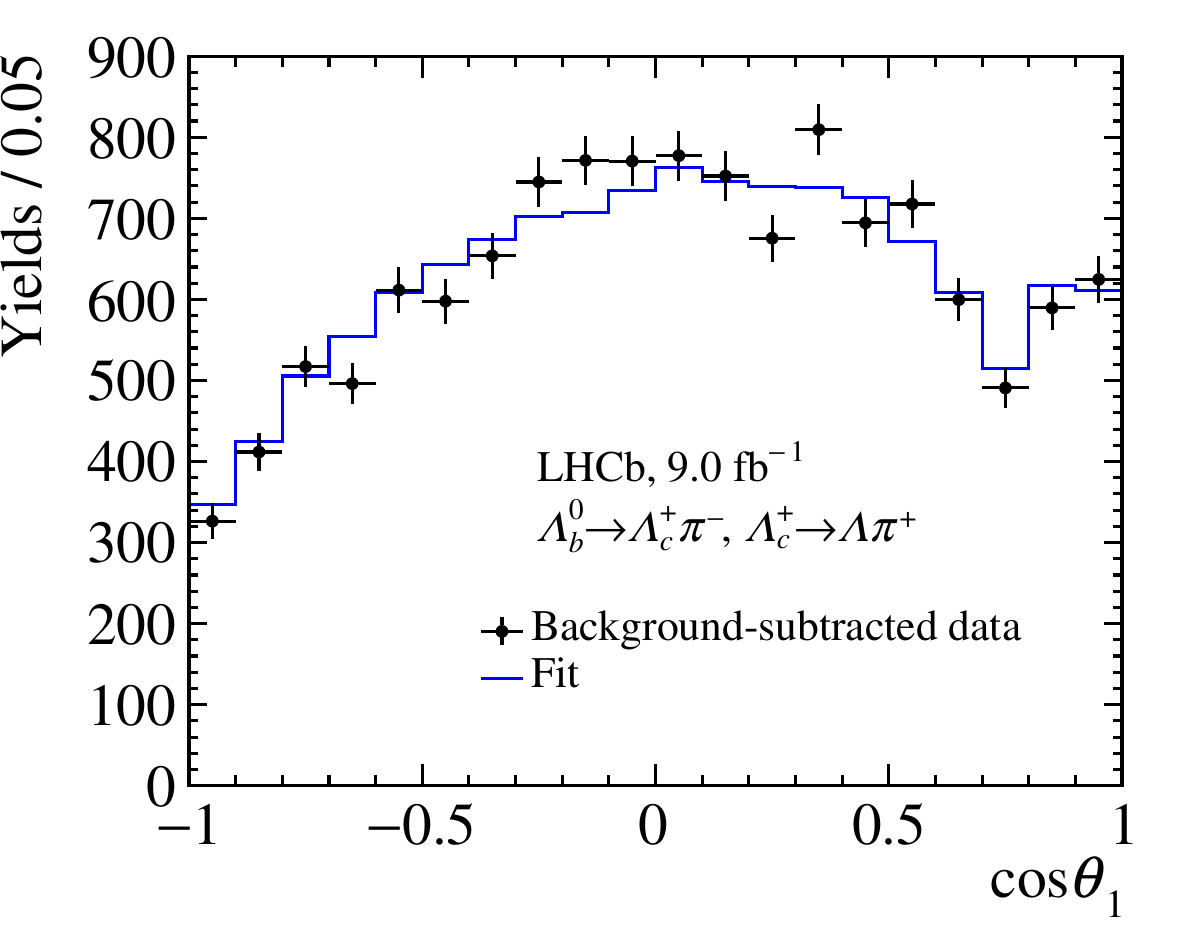}
\includegraphics[width=0.45\textwidth]{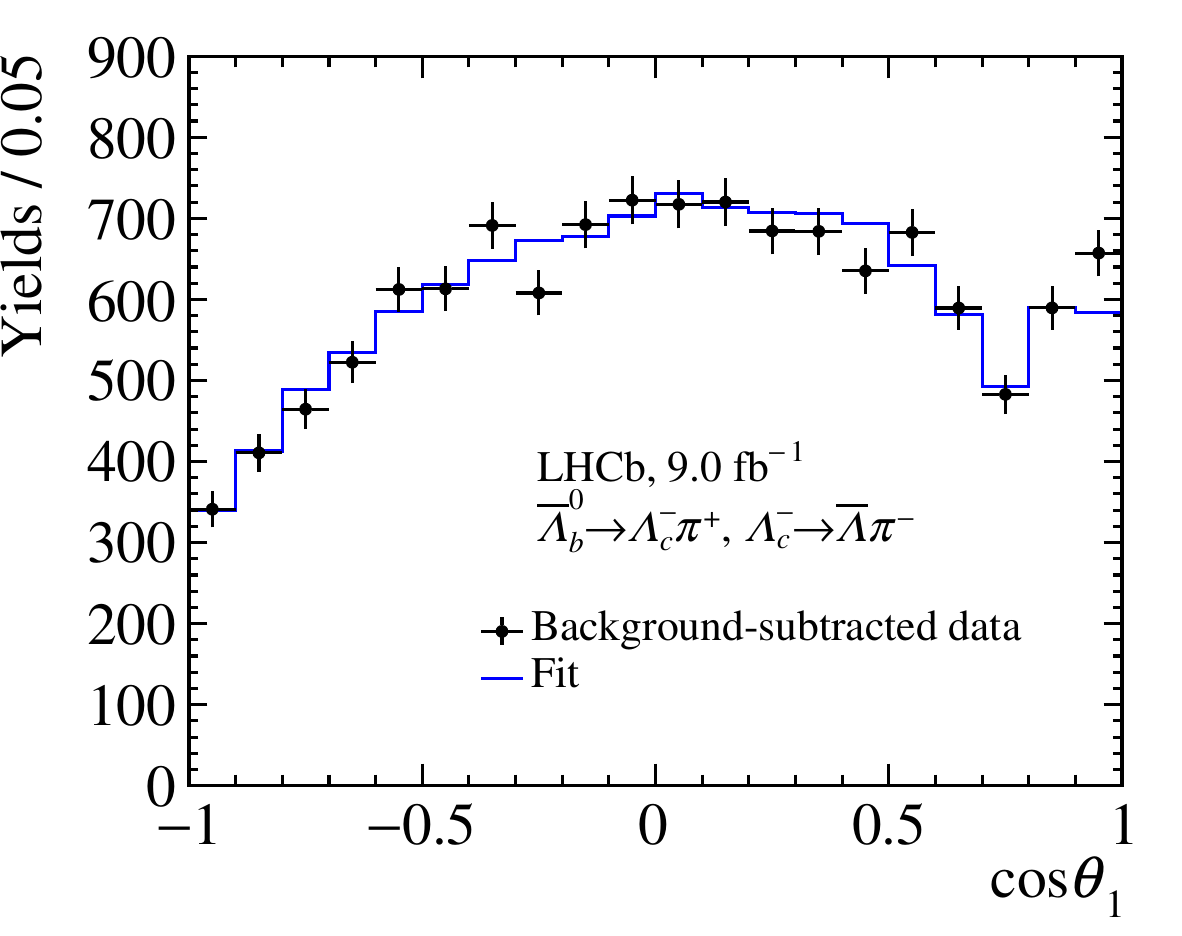}
\includegraphics[width=0.45\textwidth]{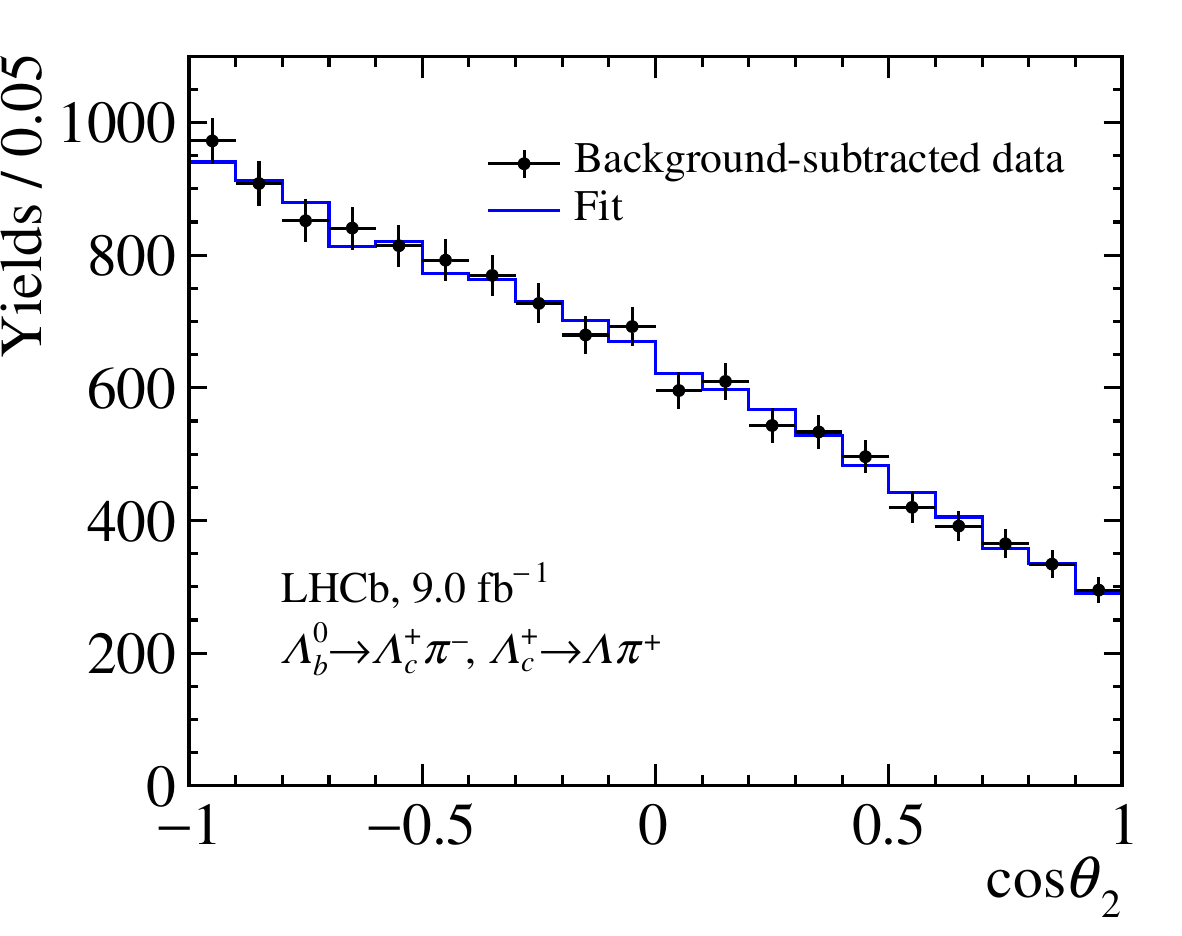}
\includegraphics[width=0.45\textwidth]{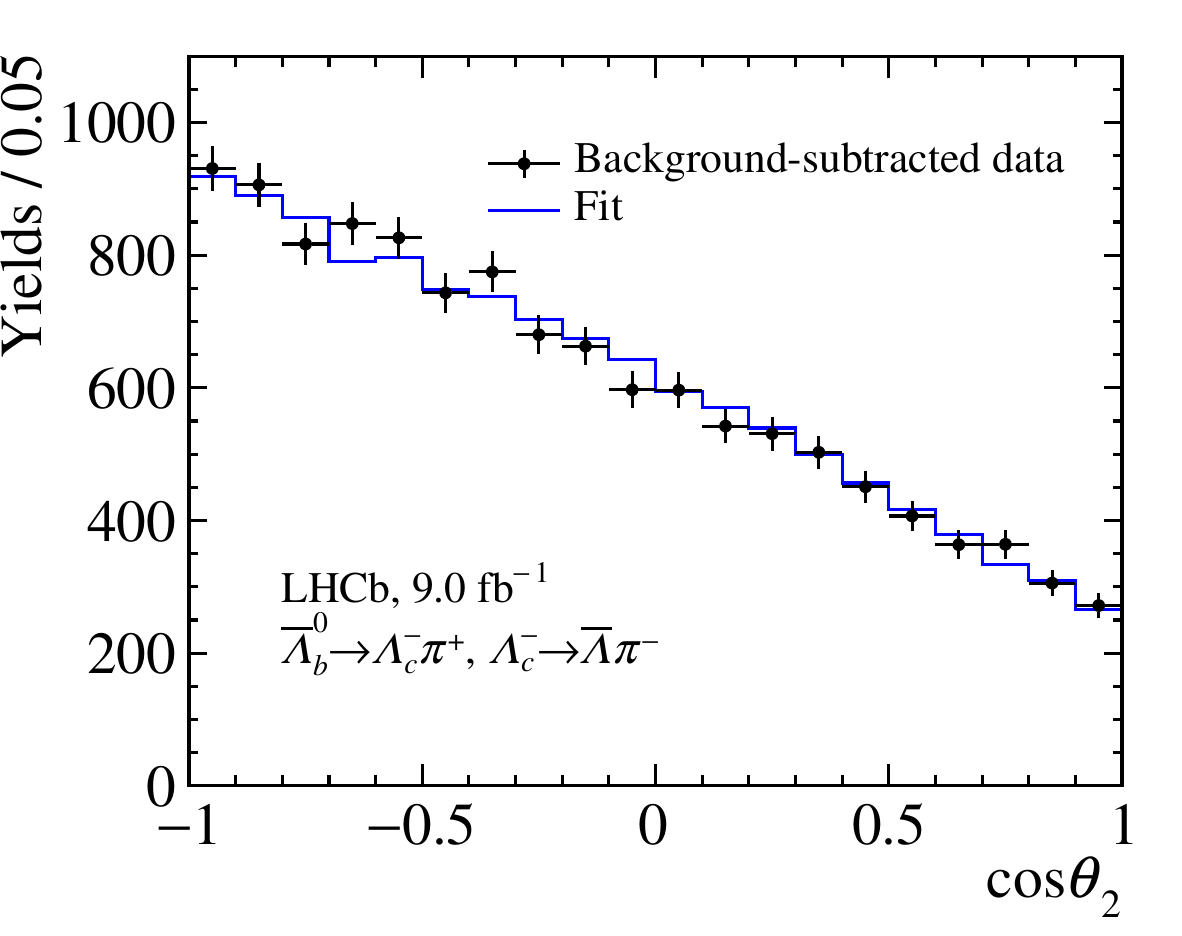}
\includegraphics[width=0.45\textwidth]{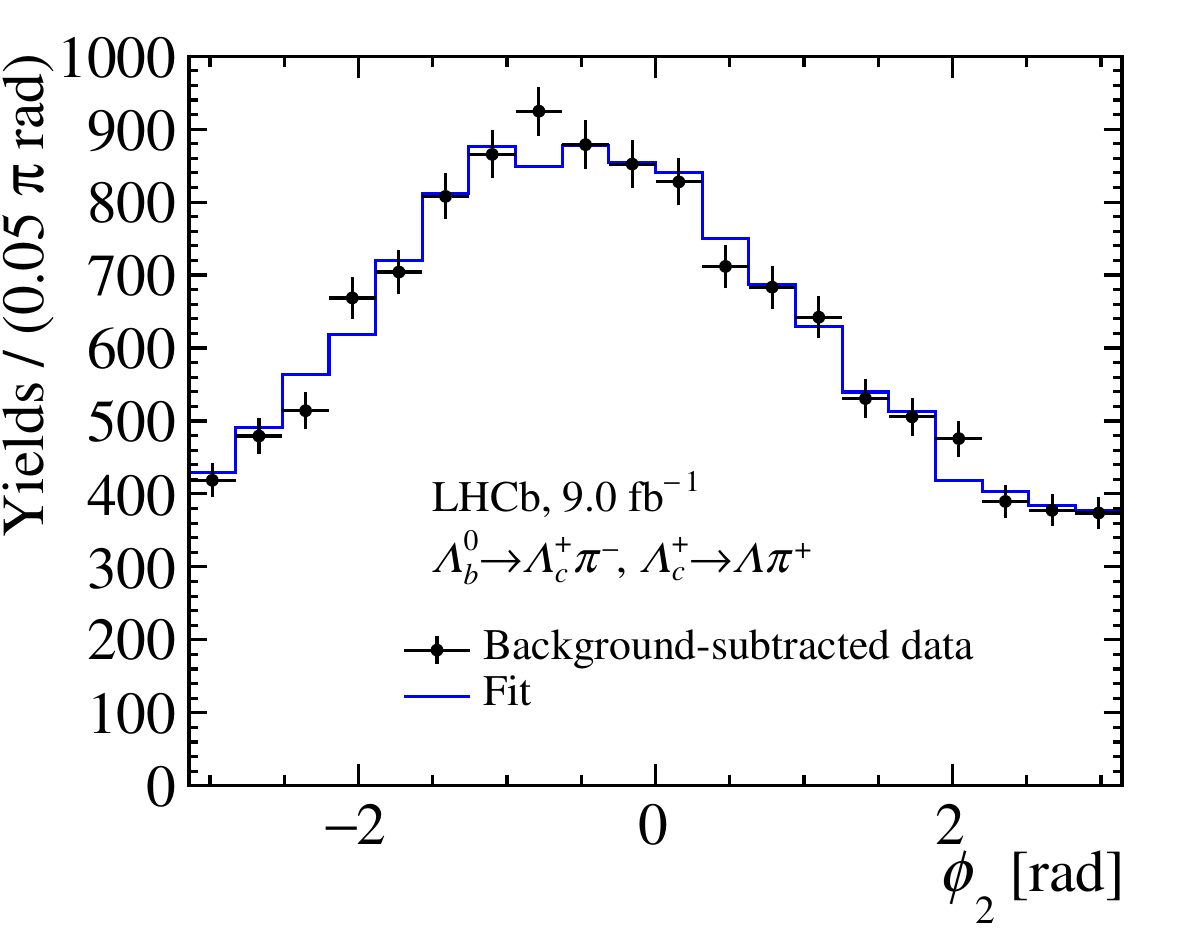}
\includegraphics[width=0.45\textwidth]{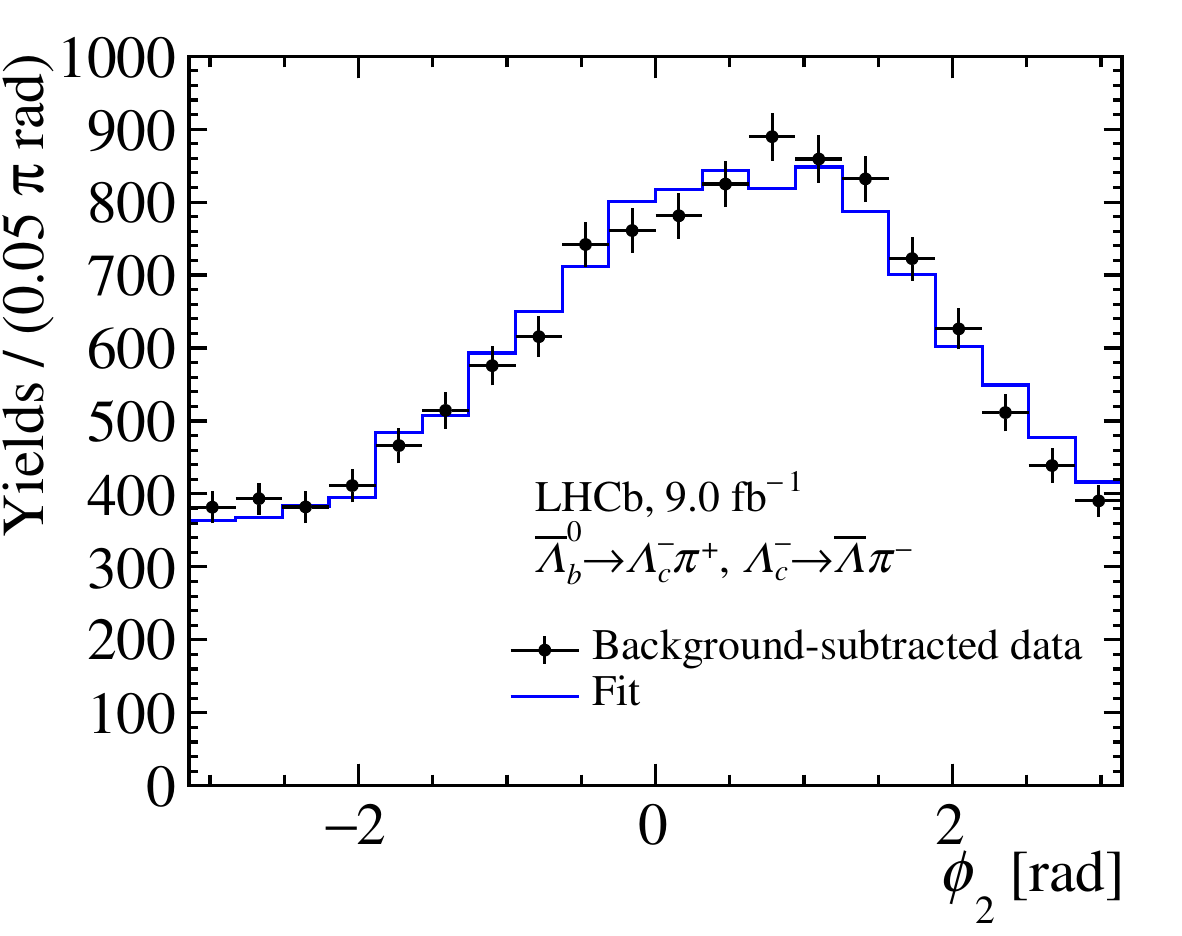}
\end{center}
\caption{Distributions of (top) $\cos\theta_1$, (middle) $\cos\theta_2$  and (bottom) $\phi_2$  for (left) the \LbToLcpiToLzpi decay  and (right) its charge-conjugate decay.
}
\label{fig:paper_Lcpi_Lzpi}
\end{figure}

\begin{figure}[htb]
\begin{center}
\includegraphics[width=0.45\textwidth]{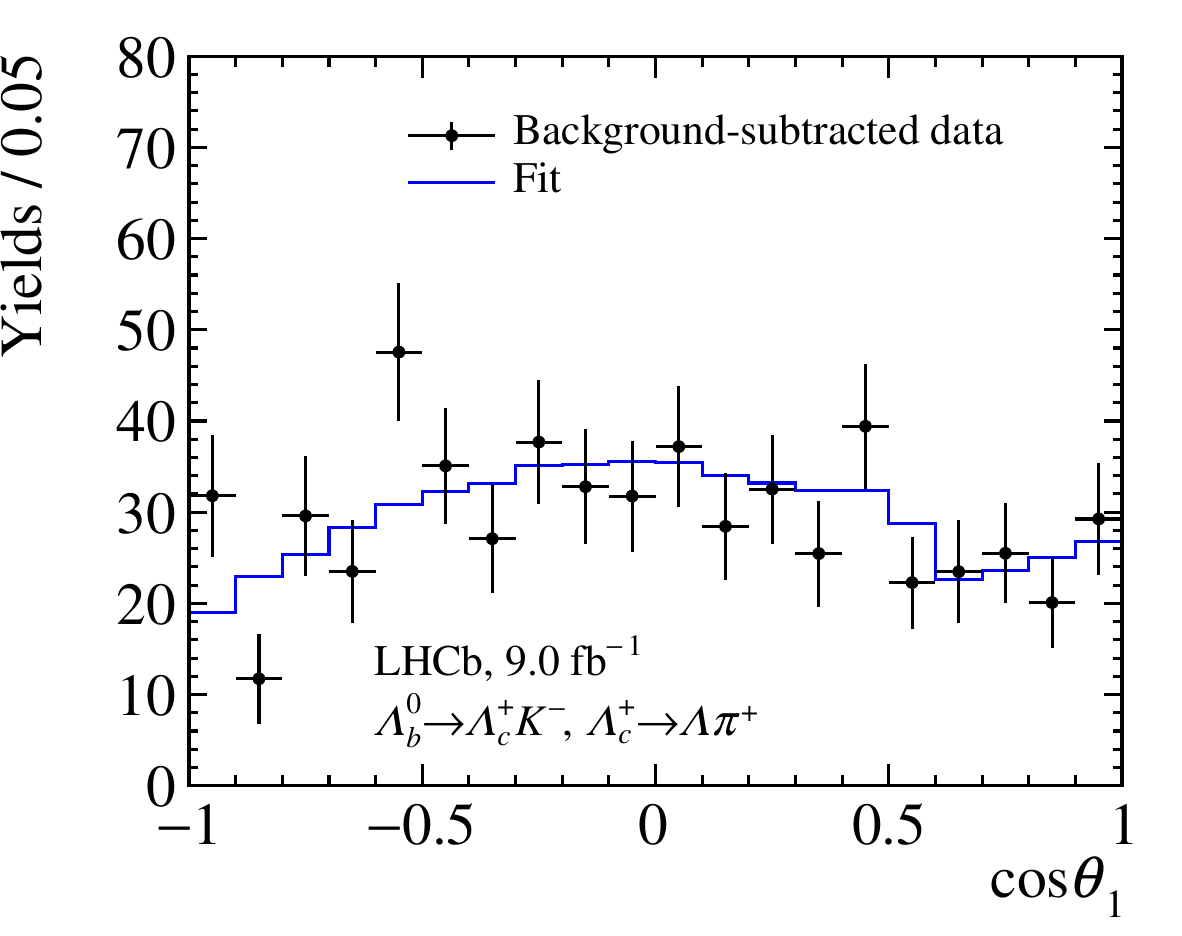}
\includegraphics[width=0.45\textwidth]{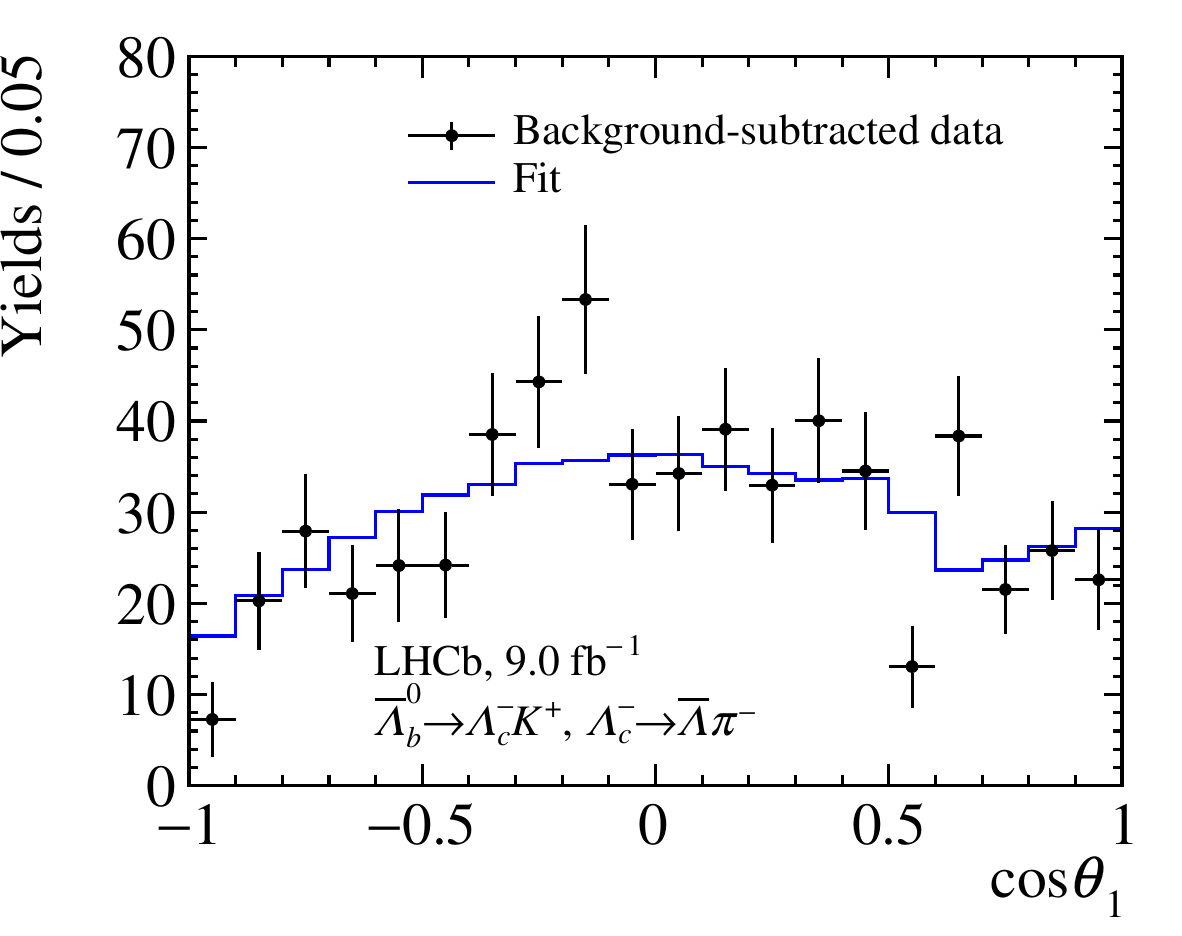}
\includegraphics[width=0.45\textwidth]{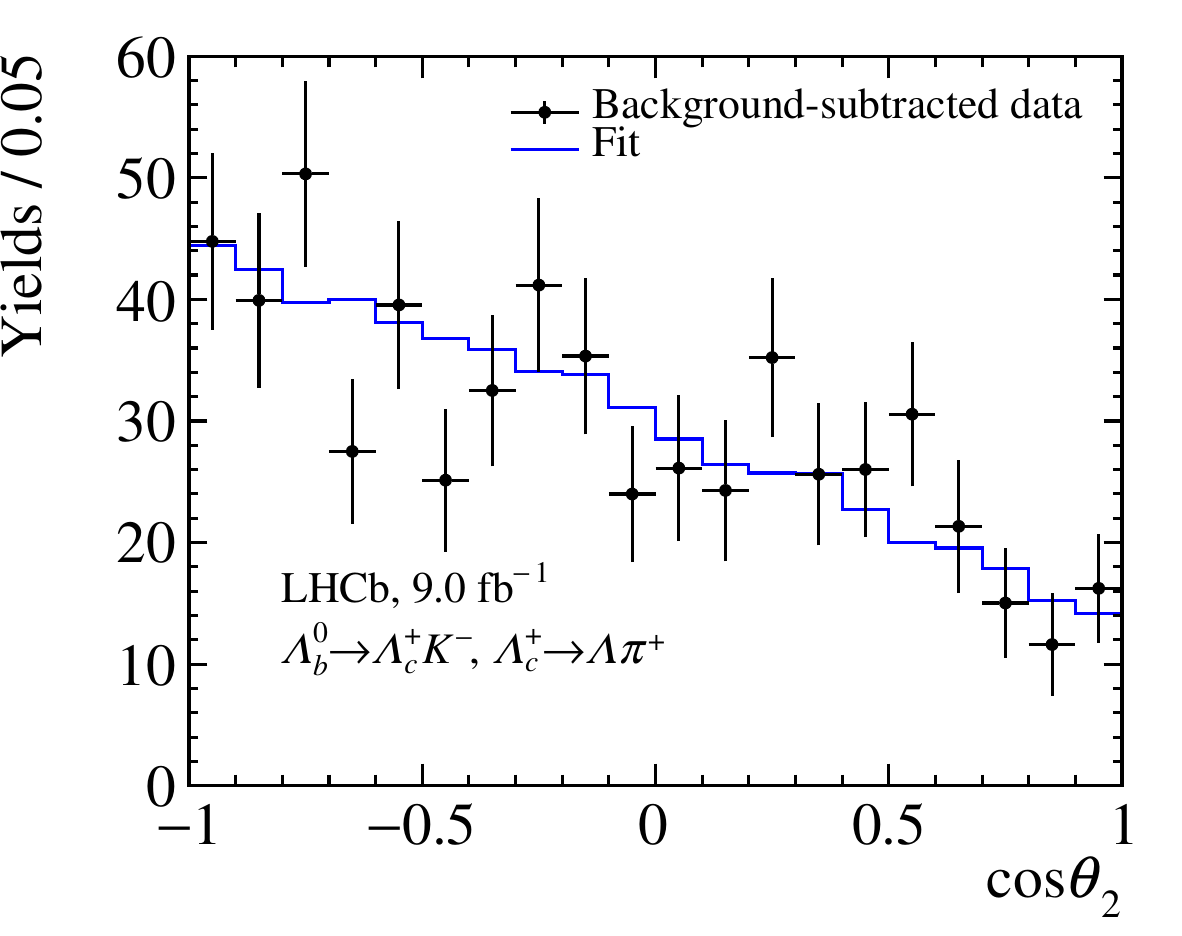}
\includegraphics[width=0.45\textwidth]{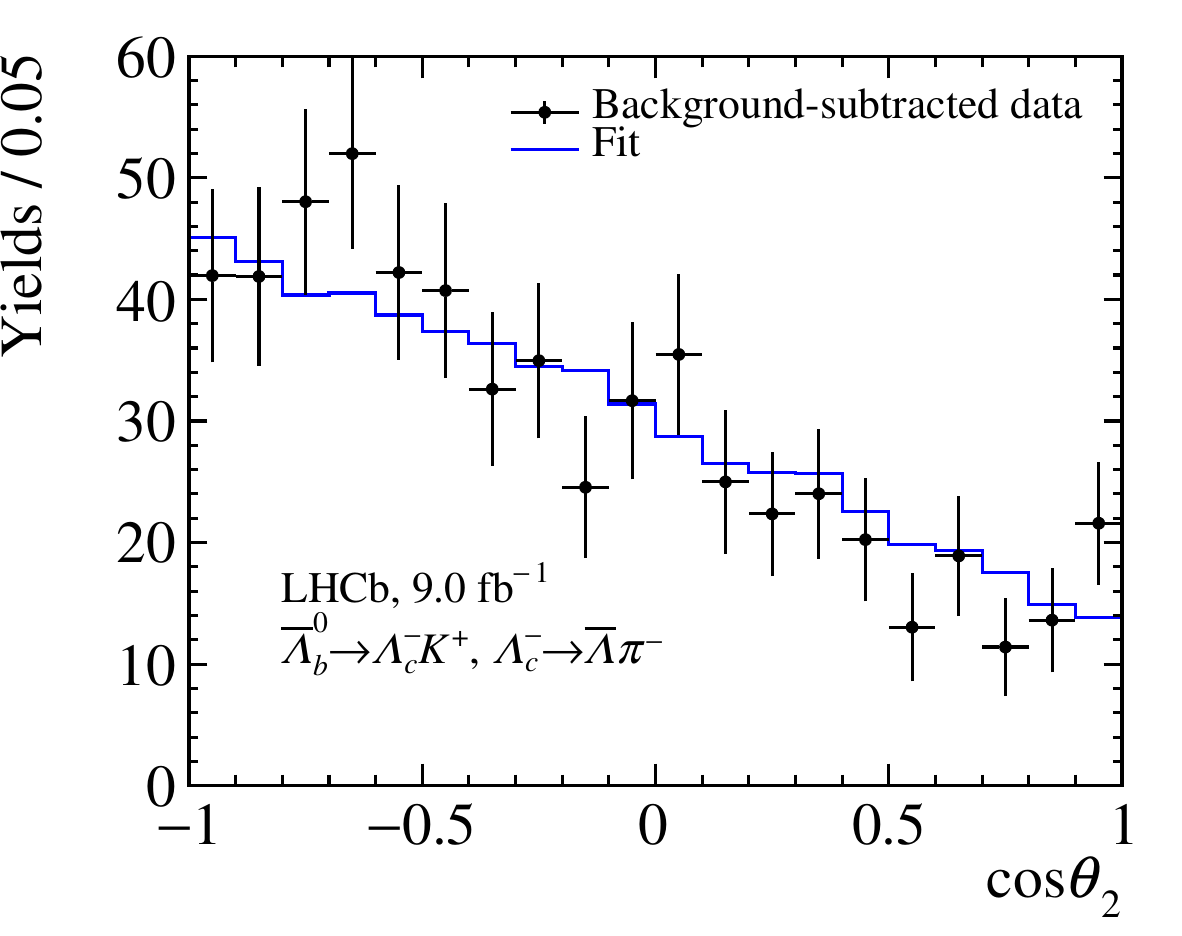}
\includegraphics[width=0.45\textwidth]{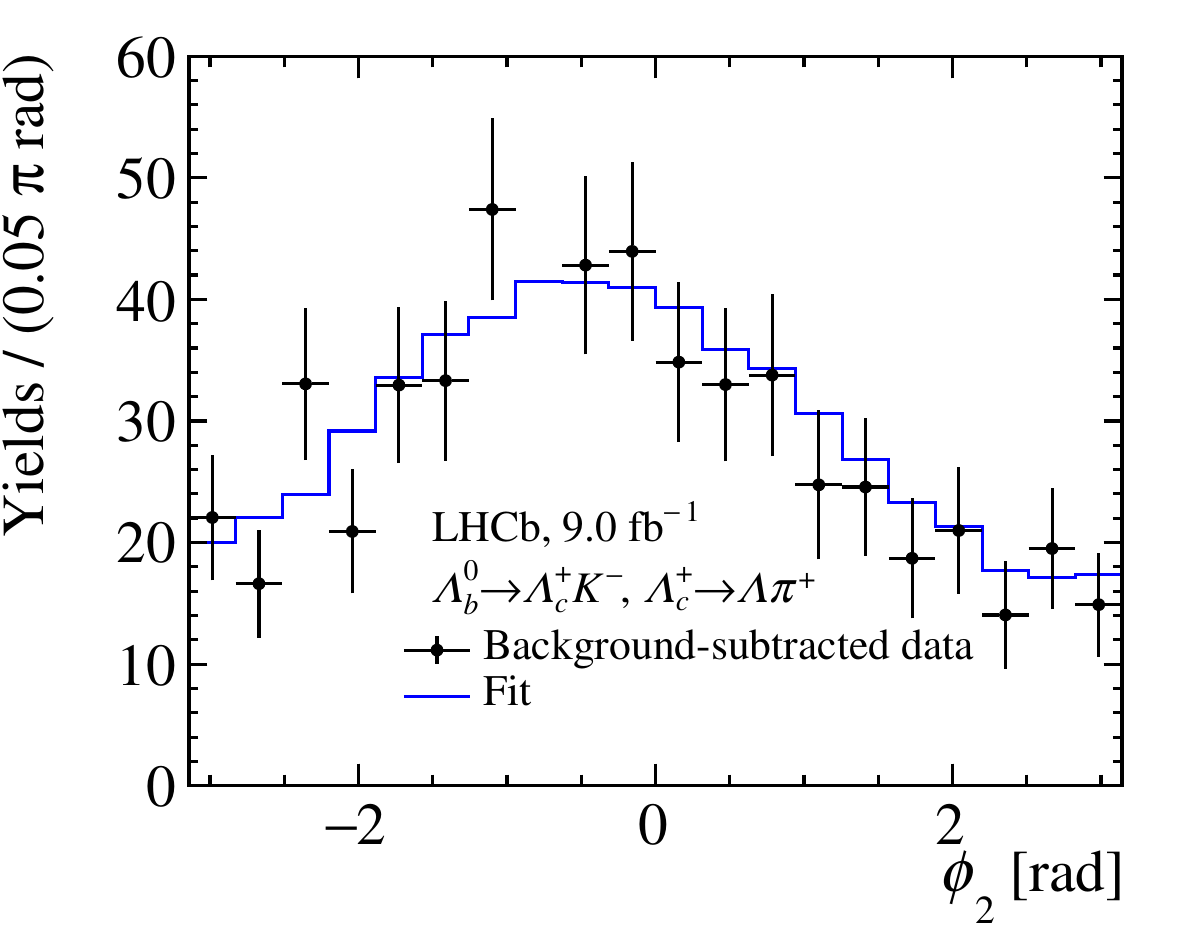}
\includegraphics[width=0.45\textwidth]{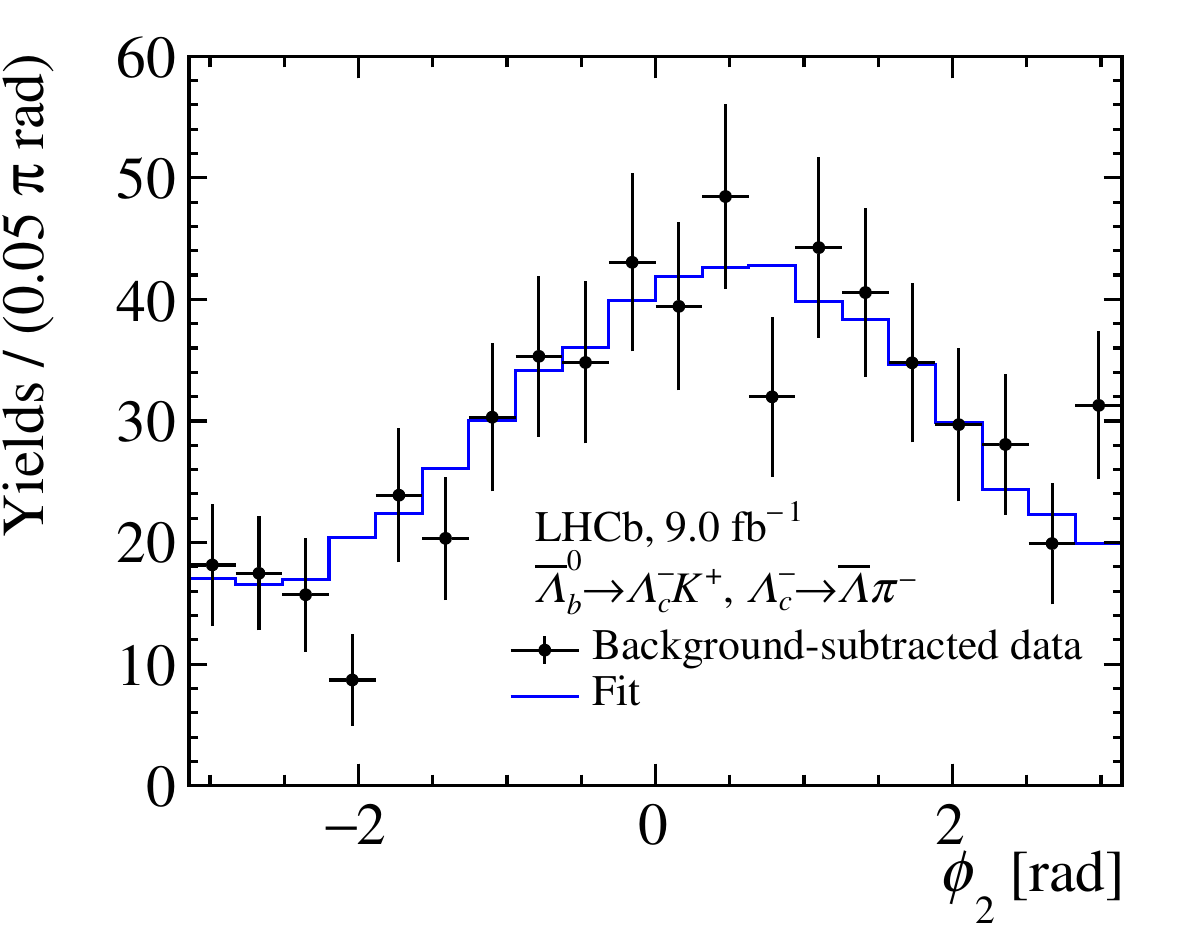}
\end{center}
\caption{Distributions of (top) $\cos\theta_1$, (middle) $\cos\theta_2$ and (bottom) $\phi_2$ for (left) the \LbToLcKToLzpi decay and (right) its charge-conjugate decay.
}
\label{fig:paper_LcK_Lzpi}
\end{figure}

\begin{figure}[htb]
\begin{center}
\includegraphics[width=0.45\textwidth]{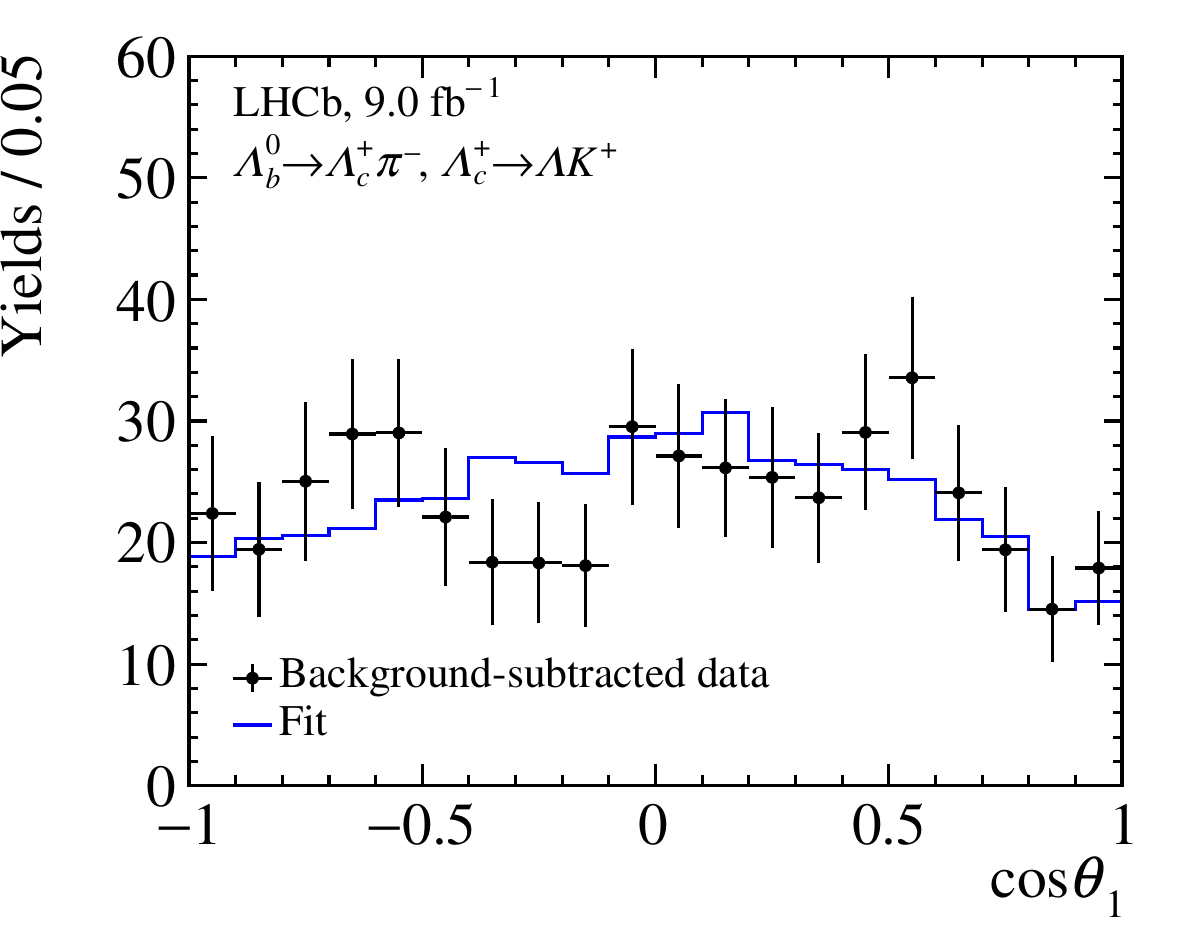}
\includegraphics[width=0.45\textwidth]{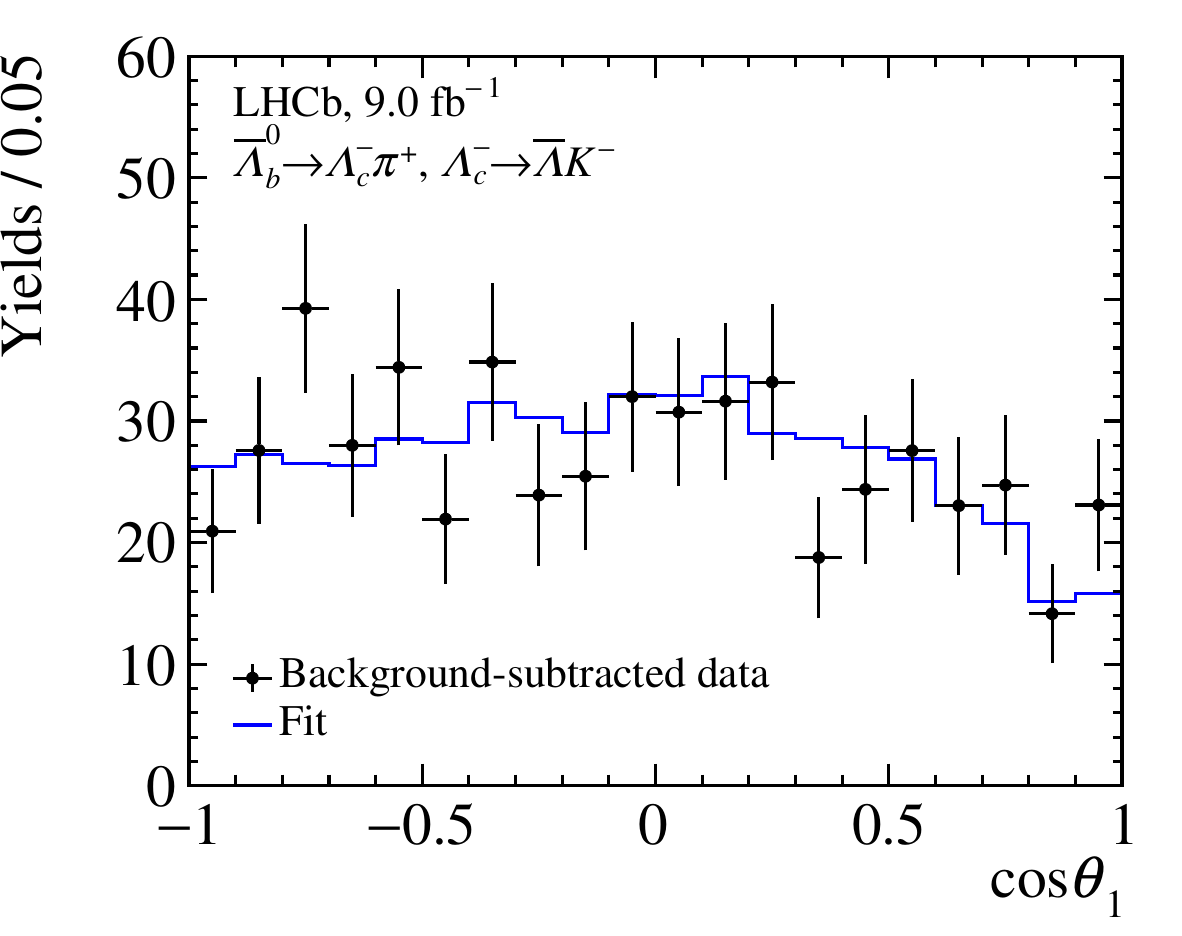}
\includegraphics[width=0.45\textwidth]{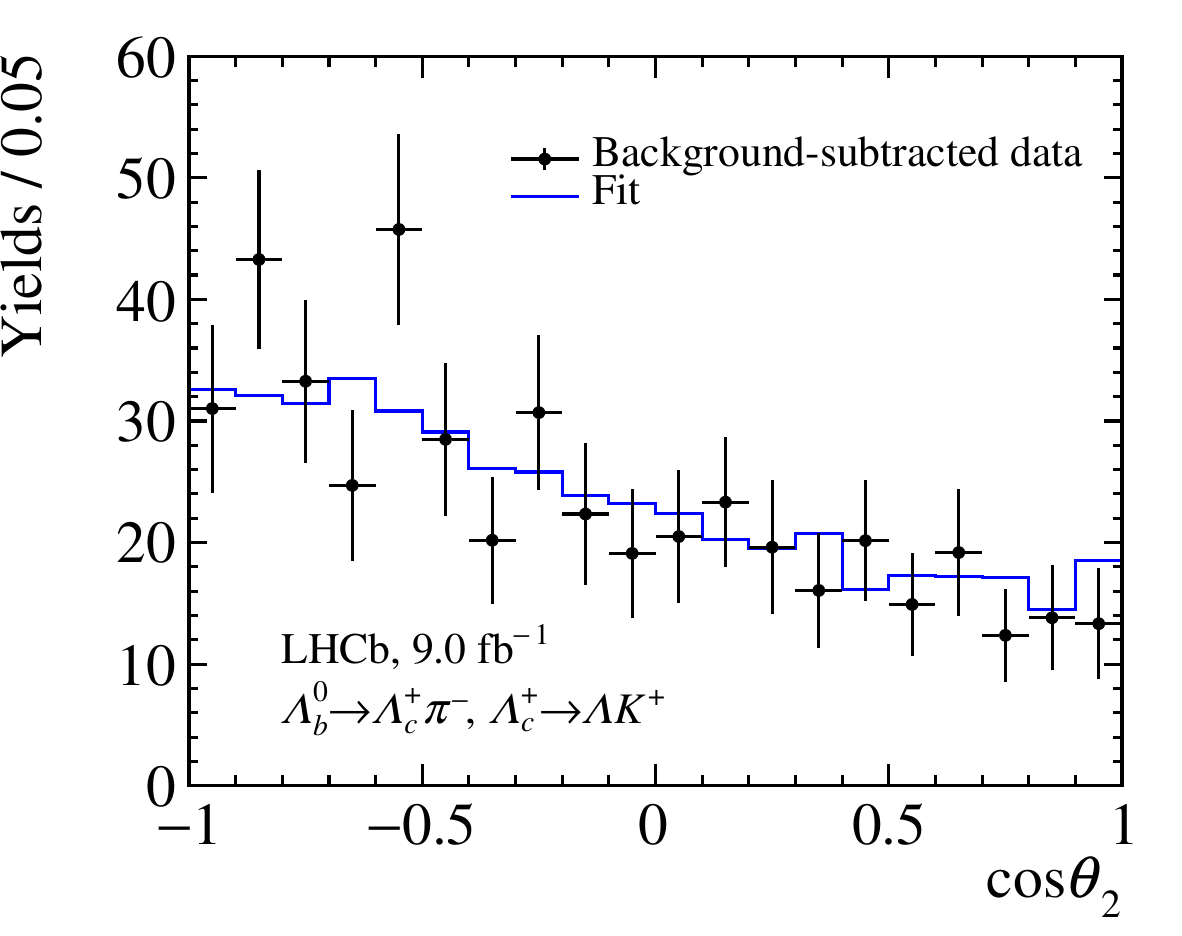}
\includegraphics[width=0.45\textwidth]{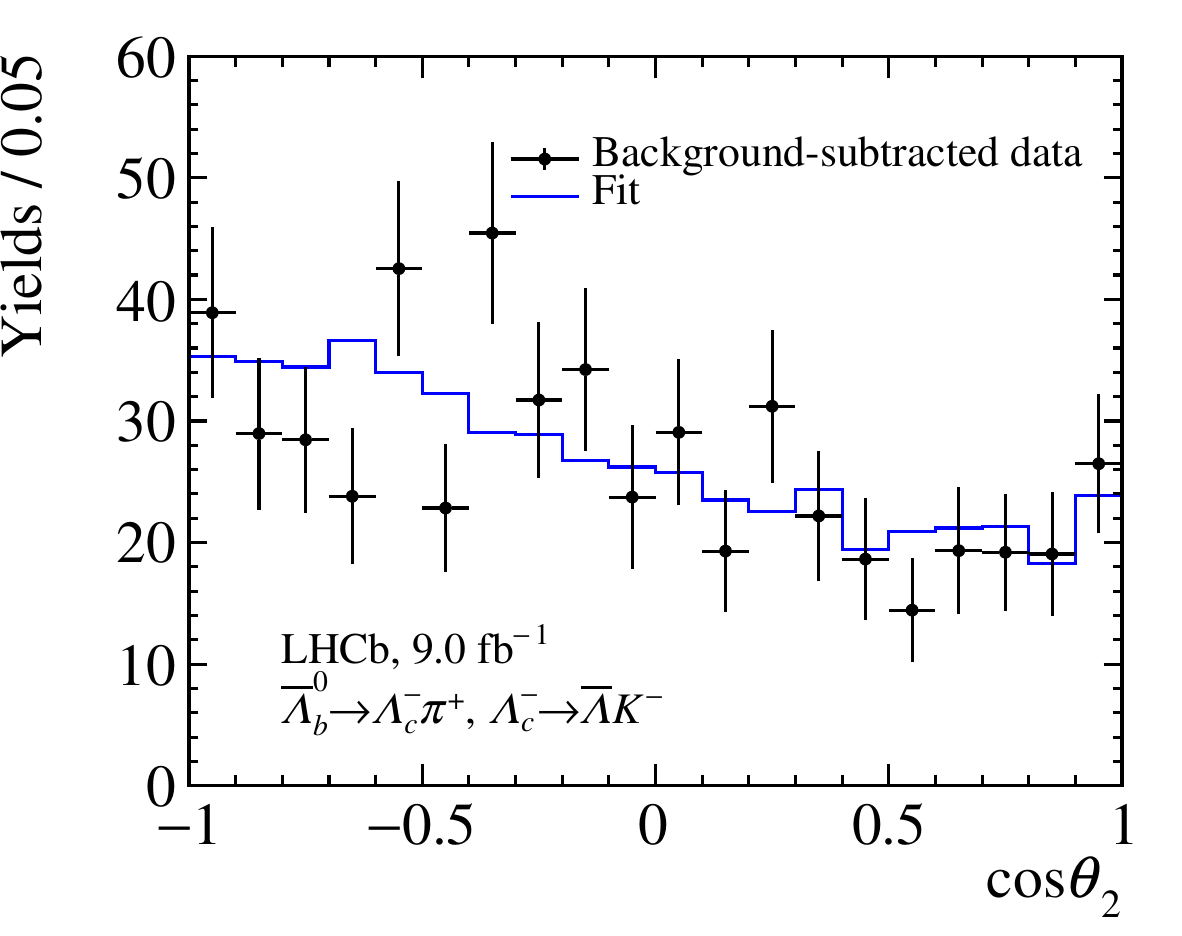}
\includegraphics[width=0.45\textwidth]{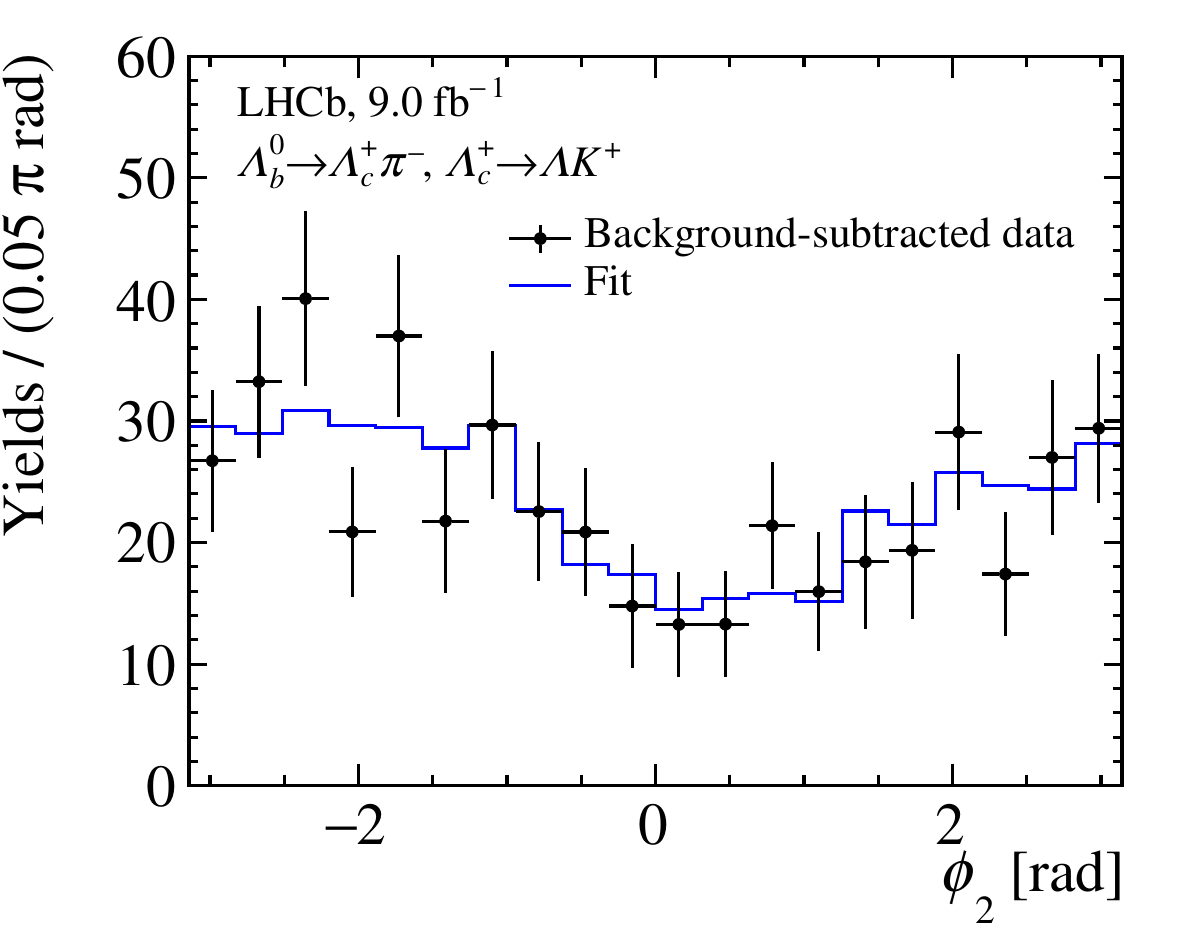}
\includegraphics[width=0.45\textwidth]{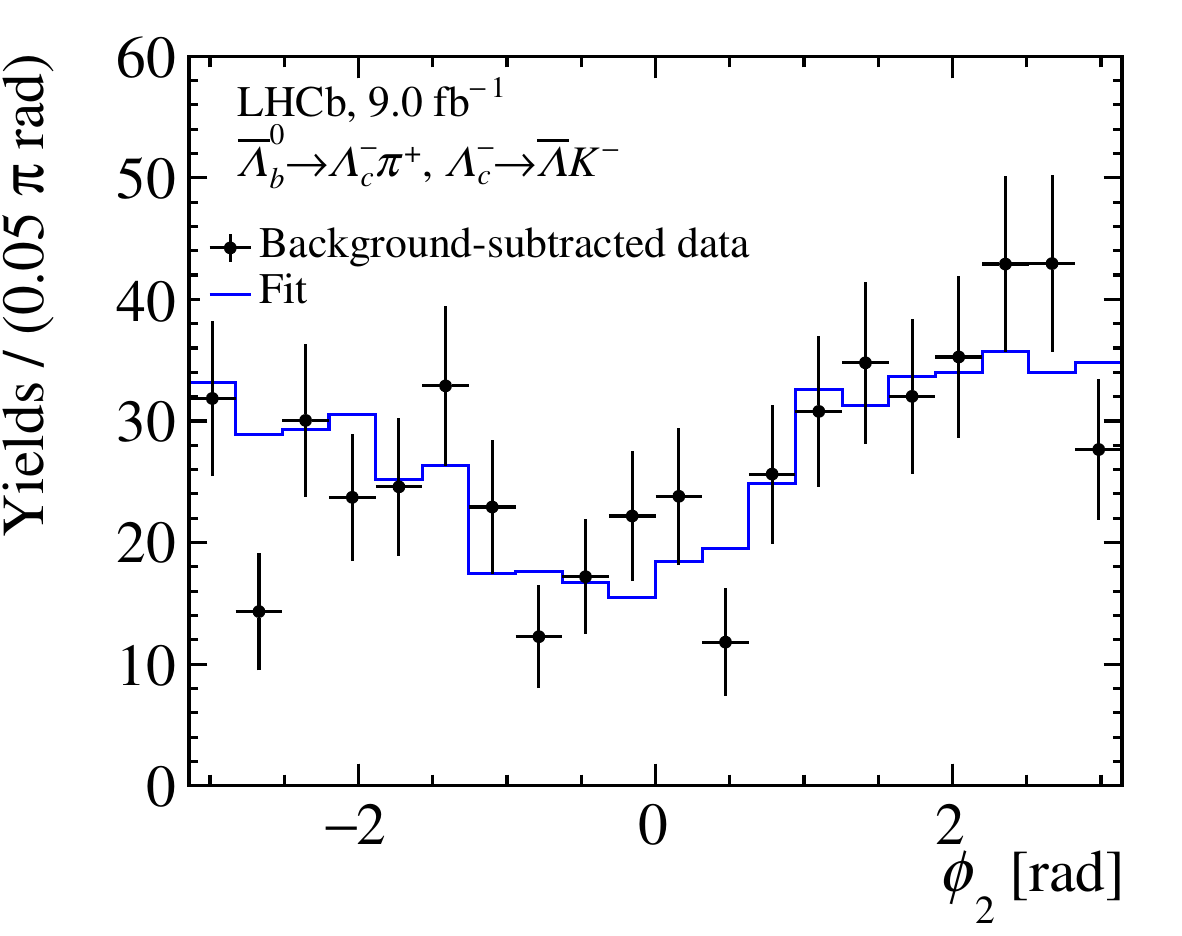}
\end{center}
\caption{Distributions of (top) $\cos\theta_1$, (middle) $\cos\theta_2$ and (bottom)  $\phi_2$  for (left) the \LbToLcpiToLzK decay  and (right) its charge-conjugate decay.
}
\label{fig:paper_Lcpi_LzK}
\end{figure}

Projections of angular distributions and fit results for various decays studied in the Letter are shown in Figs. \ref{fig:paper_Lcpi_pKS},~\ref{fig:paper_LcK_pKS},~\ref{fig:paper_Lcpi_Lzpi},~\ref{fig:paper_LcK_Lzpi} and \ref{fig:paper_Lcpi_LzK}.

\clearpage
\section{Correlation matrices for parameters}

Correlation matrices between decay parameters without accounting for systematic uncertainties are provided in Tables~\ref{tab:corr_+_1}-\ref{tab:corr_-_2}.
The correlation matrix between \CP-related parameters is provided in Table~\ref{tab:corr_cp}.

\begin{table}[htb]
\caption{Correlation matrix of decay parameters for baryon decays using $\alpha, \Delta$ as fit parameters.}
\centering
\renewcommand{\arraystretch}{1.5}
\begin{tabular}{l|rrrrrrrr}
 & \aBpi & \aBk & \aCpi & \aCk & \dCpi & \dCk & \aS & \aCp \\
\hline
\aBpi & 1.000 & 0.154 & $-$0.345 & $-$0.079 & $-$0.004 & 0.005 & 0.424 & $-$0.671 \\
\aBk & & 1.000 & $-$0.106 & $-$0.013 & 0.007 & 0.001 & 0.091 & $-$0.201 \\
\aCpi & & & 1.000 & 0.027 & 0.005 & $-$0.002 & $-$0.128 & 0.237 \\
$\alpha^{\Lambda K}_{c}$ & & & & 1.000 & $-$0.068 & $-$0.049 & $-$0.067 & 0.053 \\
\dCpi & & & & & 1.000 & $-$0.066 & 0.007 & $-$0.003 \\
\dCk & & & & & & 1.000 & 0.003 & $-$0.003 \\
\aS & & & & & & & 1.000 & $-$0.287 \\
\aCp & & & & & & & & 1.000 \\
\end{tabular}
\label{tab:corr_+_1}
\end{table}

\begin{table}[htb]
\caption{Correlation matrix of decay parameters for antibaryon decays using $\alpha, \Delta$ as fit parameters.}
\centering
\renewcommand{\arraystretch}{1.5}
\begin{tabular}{l|rrrrrrrr}
 & \baBpi & \baBk & \baCpi & \baCk & \bdCpi & \bdCk & \baS & \baCp \\
\hline
\baBpi & 1.000 & 0.164 & $-$0.306 & $-$0.035 & 0.046 & $-$0.002 & 0.380 & $-$0.660 \\
\baBk & & 1.000 & $-$0.044 & $-$0.006 & 0.021 & $-$0.001 & 0.113 & $-$0.230 \\
\baCpi & & & 1.000 & 0.011 & $-$0.021 & 0.001 & $-$0.115 & 0.201 \\
\baCk & & & & 1.000 & 0.013 & $-$0.017 & $-$0.052 & 0.023 \\
\bdCpi & & & & & 1.000 & 0.001 & $-$0.052 & $-$0.032 \\
\bdCk & & & & & & 1.000 & 0.009 & 0.001 \\
\baS & & & & & & & 1.000 & $-$0.257 \\
\baCp & & & & & & & & 1.000 \\
\end{tabular}
\label{tab:corr_-_1}
\end{table}

\begin{table}[htb]
\caption{Correlation matrix of decay parameters for baryon decays using $\beta, \gamma$ as fit parameters, with constraint $\alpha^2+\beta^2+\gamma^2=1$ imposed.}
\centering
\renewcommand{\arraystretch}{1.5}
\begin{tabular}{l|rrrrrrrr}
 & \aBpi & \aBk & \bCpi & \bCk & \gCpi & \gCk & \aS & \aCp \\
\hline
\aBpi & 1.000 & 0.154 & $-$0.120 & $-$0.016 & $-$0.189 & 0.040 & 0.424 & $-$0.671 \\
\aBk & & 1.000 & $-$0.029 & $-$0.003 & $-$0.065 & 0.006 & 0.091 & $-$0.201 \\
\bCpi & & & 1.000 & 0.002 & $-$0.592 & $-$0.005 & $-$0.106 & 0.082 \\
\bCk & & & & 1.000 & 0.002 & 0.746 & $-$0.014 & 0.011 \\
\gCpi & & & & & 1.000 & $-$0.007 & $-$0.016 & 0.130 \\
\gCk & & & & & & 1.000 & 0.021 & $-$0.027 \\
\aS & & & & & & & 1.000 & $-$0.287 \\
\aCp & & & & & & & & 1.000 \\
\end{tabular}
\label{tab:corr_+_2}
\end{table}

\begin{table}[htb]
\caption{Correlation matrix of decay parameters for antibaryon decays using $\beta, \gamma$ as fit parameters, with constraint $\alpha^2+\beta^2+\gamma^2=1$ imposed.}
\centering
\renewcommand{\arraystretch}{1.5}
\begin{tabular}{l|rrrrrrrr}
 & \baBpi & \baBk & \bbCpi & \bbCk & \bgCpi & \bgCk & \baS & \baCp \\
\hline
\baBpi & 1.000 & $-$0.164 & 0.072 & 0.002 & $-$0.204 & $-$0.021 & $-$0.380 & 0.660 \\
\baBk & & 1.000 & 0.003 & 0.000 & 0.042 & 0.021 & 0.113 & $-$0.230 \\
\bbCpi & & & 1.000 & $-$0.000 & 0.579 & $-$0.002 & $-$0.090 & 0.046 \\
\bbCk & & & & 1.000 & $-$0.001 & 0.751 & $-$0.014 & 0.011 \\
\bgCpi & & & & & 1.000 & $-$0.001 & $-$0.136 & 0.017 \\
\bgCk & & & & & & 1.000 & 0.021 & $-$0.257 \\
\baS & & & & & & & 1.000 & $-$0.257 \\
\baCp & & & & & & & & 1.000 \\
\end{tabular}
\label{tab:corr_-_2}
\end{table}

\begin{sidewaystable}
\caption{Correlation matrix for the \CP-related parameters.}
\centering
\resizebox{\textwidth}{!}{
\renewcommand{\arraystretch}{1.5}
\begin{tabular}{l|rrrrrrrrrrrrrrrr}
 & \avgBpi & \acpBpi & \avgBk & \acpBk & \avgCpi & \acpCpi & \rCpi & \rrCpi & \avgCk & \acpCk & \rCk & \rrCk & \avgCp & \acpCp & \avgS & \acpS \\
\hline
\avgBpi & 1.000 & $-$0.033 & 0.159 & $-$0.013 & $-$0.326 & 0.030 & 0.026 & $-$0.153 & $-$0.057 & 0.018 & 0.009 & $-$0.025 & $-$0.668 & 0.025 & 0.405 & 0.044 \\
\acpBpi & & 1.000 & $-$0.007 & 0.158 & $-$0.026 & $-$0.326 & $-$0.097 & $-$0.031 & $-$0.022 & $-$0.054 & $-$0.010 & $-$0.014 & $-$0.011 & $-$0.666 & $-$0.032 & $-$0.404 \\
\avgBk & & & 1.000 & $-$0.164 & $-$0.077 & 0.036 & $-$0.017 & $-$0.029 & $-$0.004 & $-$0.009 & $-$0.002 & $-$0.004 & $-$0.214 & 0.001 & $-$0.101 & $-$0.001 \\
\acpBk & & & & 1.000 & $-$0.039 & $-$0.078 & $-$0.015 & $-$0.024 & $-$0.004 & $-$0.009 & $-$0.002 & $-$0.003 & $-$0.005 & $-$0.213 & $-$0.000 & $-$0.100 \\
\avgCpi & & & & & 1.000 & $-$0.023 & $-$0.020 & $-$0.531 & 0.019 & $-$0.007 & $-$0.003 & $-$0.005 & 0.220 & $-$0.022 & $-$0.122 & $-$0.012 \\
\acpCpi & & & & & & 1.000 & 0.367 & 0.005 & 0.018 & 0.008 & 0.003 & 0.005 & 0.019 & 0.219 & 0.009 & 0.122 \\
\rCpi & & & & & & & 1.000 & 0.043 & $-$0.010 & 0.004 & 0.002 & 0.004 & $-$0.102 & $-$0.015 & 0.065 & 0.010 \\
\rrCpi & & & & & & & & 1.000 & $-$0.010 & 0.001 & 0.002 & 0.003 & $-$0.018 & 0.023 & 0.065 & 0.015 \\
\avgCk & & & & & & & & & 1.000 & 0.095 & $-$0.077 & $-$0.428 & 0.038 & $-$0.015 & $-$0.033 & $-$0.017 \\
\acpCk & & & & & & & & & & 1.000 & 0.147 & 0.025 & 0.011 & 0.036 & 0.013 & 0.032 \\
\rCk & & & & & & & & & & & 1.000 & 0.188 & 0.006 & 0.036 & 0.010 & 0.010 \\
\rrCk & & & & & & & & & & & & 1.000 & $-$0.017 & 0.009 & 0.013 & 0.015 \\
\avgCp & & & & & & & & & & & & & 1.000 & $-$0.005 & $-$0.275 & $-$0.023 \\
\acpCp & & & & & & & & & & & & & & 1.000 & 0.021 & 0.273 \\
\avgS & & & & & & & & & & & & & & & 1.000 & 0.046 \\
\acpS & & & & & & & & & & & & & & & & 1.000 \\
\end{tabular}
}
\label{tab:corr_cp}
\end{sidewaystable}

\clearpage

\newpage
\centerline
{\large\bf LHCb collaboration}
\begin
{flushleft}
\small
R.~Aaij$^{36}$\lhcborcid{0000-0003-0533-1952},
A.S.W.~Abdelmotteleb$^{55}$\lhcborcid{0000-0001-7905-0542},
C.~Abellan~Beteta$^{49}$,
F.~Abudin{\'e}n$^{55}$\lhcborcid{0000-0002-6737-3528},
T.~Ackernley$^{59}$\lhcborcid{0000-0002-5951-3498},
A. A. ~Adefisoye$^{67}$\lhcborcid{0000-0003-2448-1550},
B.~Adeva$^{45}$\lhcborcid{0000-0001-9756-3712},
M.~Adinolfi$^{53}$\lhcborcid{0000-0002-1326-1264},
P.~Adlarson$^{80}$\lhcborcid{0000-0001-6280-3851},
C.~Agapopoulou$^{13}$\lhcborcid{0000-0002-2368-0147}
% ...

{\footnotesize \it

% ...
$^{13}$Universit{\'e} Paris-Saclay, CNRS/IN2P3, IJCLab, Orsay, France\\
% ...
$^{36}$Nikhef National Institute for Subatomic Physics, Amsterdam, Netherlands\\
% ...
$^{45}$Instituto Galego de F{\'\i}sica de Altas Enerx{\'\i}as (IGFAE), Universidade de Santiago de Compostela, Santiago de Compostela, Spain\\
% ...
$^{49}$Physik-Institut, Universit{\"a}t Z{\"u}rich, Z{\"u}rich, Switzerland\\
% ...
$^{53}$H.H. Wills Physics Laboratory, University of Bristol, Bristol, United Kingdom\\
% ...
$^{55}$Department of Physics, University of Warwick, Coventry, United Kingdom\\
% ...
$^{58}$School of Physics and Astronomy, University of Glasgow, Glasgow, United Kingdom\\
$^{59}$Oliver Lodge Laboratory, University of Liverpool, Liverpool, United Kingdom\\
% ...
$^{67}$Syracuse University, Syracuse, NY, United States\\
% ...
$^{80}$Department of Physics and Astronomy, Uppsala University, Uppsala, Sweden, associated to $^{58}$\\
% ...
}
\end{flushleft}

\end{document}